\documentclass[runningheads, a4paper]{llncs}
\usepackage[english]{babel}

\usepackage{amsmath, amssymb, amsthm}
\usepackage{graphicx}
\usepackage{stmaryrd}
\usepackage[T1]{fontenc}
\usepackage[scaled]{beramono}
\usepackage{listings}
\usepackage{color}
\usepackage[utf8]{inputenc}
\usepackage{babel}
\usepackage{times}
\usepackage{mathpartir}
\usepackage{boxedminipage}
\usepackage[]{algorithm2e}
\usepackage{subfiles}
\usepackage{tikz}
\usepackage{subcaption}
\usepackage{prftree}
\usepackage[scaled=0.9]{newtxtt} 
\usepackage{xspace}
\usepackage{mathtools}

\tikzset{nd/.style={rectangle,draw,fill=gray!20,rounded corners=1mm,inner ysep=1.5pt,inner xsep=4pt}}

\AtBeginDocument{}
\AtBeginDocument{\normalsize}

\makeatletter

\renewcommand\section{\@startsection{section}{1}{\z@}{-14\p@ \@plus -4\p@ \@minus -4\p@}{9\p@ \@plus 4\p@ \@minus 4\p@}{\normalfont\large\bfseries\boldmath\rightskip=\z@ \@plus 8em\pretolerance=10000 }}
\renewcommand\subsection{\@startsection{subsection}{1}{\z@}{-15\p@ \@plus -4\p@ \@minus -4\p@}{9\p@ \@plus 4\p@ \@minus 4\p@}{\normalfont\normalsize\bfseries\boldmath\rightskip=\z@ \@plus 8em\pretolerance=10000 }}

\newcommand{\boxedsymbol}[2]{%
  \begingroup
  \setlength{\fboxsep}{0pt}%
  \setlength{\fboxrule}{\variable@rule{#1}}%
  \fbox{%
    $\m@th#1\variable@space{#1}#2\variable@space{#1}$%
  }%
  \endgroup
}
\newcommand{\variable@rule}[1]{%
  \fontdimen8  
  \ifx#1\displaystyle\textfont3\else
    \ifx#1\textstyle\textfont3\else
      \ifx#1\scriptstyle\scriptfont3\else
        \scriptscriptfont3\relax
  \fi\fi\fi
}
\newcommand{\variable@space}[1]{%
  \mspace{-
    \ifx#1\displaystyle 1.25\else
      \ifx#1\textstyle 1.25\else
        \ifx#1\scriptstyle 1.4\else
          1.6%
    \fi\fi\fi
  mu}%
}
\makeatother

\newcommand{\boxbowtie}{\mathbin{\mathpalette\boxedsymbol\bowtie}}

\newcommand{\vars}[1]{\ensuremath{\mathit{vars}(#1)}}
\newcommand{\dom}[1]{\ensuremath{\mathit{dom}(#1)}}

\definecolor{bluekeywords}{rgb}{0.13,0.13,1}
\definecolor{greencomments}{rgb}{0,0.5,0}
\definecolor{redstrings}{rgb}{0.9,0,0}
\definecolor{cyantypes}{rgb}{0.18,0.56,0.72}
\lstdefinelanguage{lambda}%
{morekeywords={let, new, match, with, rec, open, module, namespace, of, member, %
and, for, while, true, false, in, do, begin, end, fun, function, return, yield, try, %
mutable, if, then, else, cloud, async, static, use, abstract, interface,
inherit, finally, type, shape },
  otherkeywords={ let!, return!, do!, yield!, use!, var, from, select, where,
  order, by, query, head },
  keywordstyle=\color{bluekeywords},
  sensitive=true,
  basicstyle=\ttfamily,
  breaklines=true,
  xleftmargin=\parindent,
  tabsize=4,
  morecomment=[l][\color{greencomments}]{///},
  morecomment=[l][\color{greencomments}]{//},
  morecomment=[s][\color{greencomments}]{{(*}{*)}},
  morestring=[b]",
  showstringspaces=false,
  literate={`}{\`}1,
  stringstyle=\color{redstrings},
  classoffset=1,
  morekeywords={int, string, MyNewType, hasAge, hasName},
  keywordstyle=\color{cyantypes},
  classoffset=0,
  mathescape=true,
  numbers=left
}

\newif\ifpaper

\newcommand{\ifShort}[2]{%
\ifpaper%
#1%
\else%
#2%
\fi
}%

\newcommand\myeq{\stackrel{\mathclap{\normalfont\mbox{def}}}{=}}

\begin{document}


\ifShort{\title{Type Checking Program Code using SHACL}}
    {\title{Type Checking Program Code using SHACL (Extended Version)}}
 
\author{
	Martin Leinberger\inst{1}, 
  Philipp Seifer\inst{2},
  Claudia Schon\inst{1},
	Ralf Lämmel\inst{2}, 
	Steffen Staab\inst{1,3}
}

\institute{
    $^1$ Institute \text{for} Web Science and Technologies, University of
    Koblenz-Landau, Germany \\ 
    $^2$ The Software Languages Team, University of Koblenz-Landau, Germany \\ 
    $^3$ Web and Internet Science Research Group, University of Southampton,
    England 
}

\maketitle

\begin{abstract} 
It is a strength of graph-based data formats, like RDF, that they are very
flexible with representing data. To avoid run-time errors, program code that
processes highly-flexible data representations exhibits the difficulty that it
must always include the most general case, in which attributes might be
set-valued or possibly not available. The Shapes Constraint Language (SHACL) has
been devised to enforce constraints on otherwise random data structures. We
present our approach, Type checking using SHACL (TyCuS), for type checking code
that queries RDF data graphs validated by a SHACL shape graph.  To this end, we
derive SHACL shapes from queries and integrate data shapes and query shapes as
types into a $\lambda$-calculus.  We provide the formal underpinnings and
a proof of type safety for TyCuS. A programmer can use our method in order to
process RDF data with simplified, type checked code that will not encounter
run-time errors (with usual exceptions 
as type checking cannot prevent accessing
empty lists).
\end{abstract}

\section{Introduction}
\label{sec:introduction} 

Graph-based data formats, such as RDF, have become increasingly popular, because
they allow for much more flexibility for describing data items than
rigidly-structured relational databases. Even when an ontology defines classes
and properties, because of its open-world assumption, it is always possible to
leave away required information or to add new classes and properties on the fly.
Such flexibility incurs cost. 
%
Programmers cannot rely on structural restrictions of data
relationships.
%
For instance, the
following T-Box axiom states that every \texttt{Student} has at least one
\texttt{studiesAt} relation:
\begin{equation}
\texttt{Student} \sqsubseteq \mathop{\geq 1}\,\texttt{studiesAt}.\top 
\end{equation}
Consider an RDF data graph such as shown in Fig.~\ref{fig:ex:g1}. The two nodes
\texttt{alice} and \texttt{bob} are both instances of \texttt{Student} and
\texttt{Person}. For \texttt{alice}, only the name is known. For \texttt{bob},
name, age and that he studies at $b_1$, which is an instance of
\texttt{University}.  
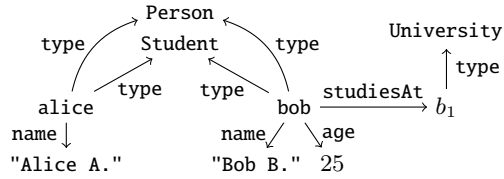
\begin{figure}[tb]
\centering
\begin{tikzpicture}[auto,scale=1.0,transform shape]
    \begin{scope}[every node/.style={font=\ttfamily\footnotesize}]
        \node (alice) at (-1.5,0) {alice};
        \node (aliceName) at (-1.5,-0.75) {"Alice A."};
        \node (bob) at (1.5,0) {bob};
        \node (bobName) at (1.0,-0.75) {"Bob B."};
        \node (bobAge) at (2.0,-0.75) {$25$};
        \node (b1) at (3.5,0) {$b_1$};

        \node (stud) at (0,0.85) {Student};
        \node (pers) at (0,1.25) {Person}; 
        \node (uni)  at (3.5,1) {University};
    \end{scope}

    \begin{scope}[every edge/.style={draw=black},
        every node/.style={font=\ttfamily\footnotesize}] 
        \draw [->] (alice) edge[right] node[pos=0.25,yshift=-0.1cm]{type} (stud);
        \draw [->] (alice) edge node[left] {name} (aliceName);
        \draw [->] (alice) edge[left,bend left] node{type} (pers);
        \draw [->] (bob) edge[left] node[pos=0.25,yshift=-0.1cm]{type} (stud);
        \draw [->] (bob) edge[right,bend right] node{type} (pers);
        \draw [->] (bob) edge node[left] {name} (bobName);
        \draw [->] (bob) edge node[right] {age} (bobAge);
        \draw [->] (b1) edge[right] node{type} (uni);
        \draw [->] (bob) edge[above] node{studiesAt} (b1);
    \end{scope}
\end{tikzpicture}    
\caption{Sample RDF data graph $G_1$.}
\label{fig:ex:g1}
\end{figure}
Such a graph is a valid A-Box for the T-Box stated above. However, for a program
containing a variable $x$ representing an instance of
\texttt{Student}, there is no guarantee that the place of study is explicitly
mentioned in the data and can be displayed. Depending on whether $x$ contains
\texttt{alice} or \texttt{bob}, the following program may succeed or encounter
a run-time error:
\begin{lstlisting}[language=lambda]
print(x.studiesAt)
\end{lstlisting}
The Shapes Constraint Language (SHACL) is a recent W3C
recommendation~\cite{SHACL} set out to allow for formulating integrity
constraints. By now, a proposal for its formal semantics has been formulated by
the research community~\cite{SHACL-Semantics} and SHACL shape graphs can be used
to validate given data graphs.
\cite{SHACL} itself states that:
\begin{quote}
   SHACL shape graphs [...]
    may be used for a variety of purposes besides validation, including user
    interface building, code generation and data integration. 
\end{quote}
However, it does not state \emph{how} SHACL shape graphs might be used for these
purposes. 
We consider the problem of writing code against an---possibly evolving---RDF
data graph that is and remains conformant to a SHACL shape graph.  We assume
that the RDF database handles the rejection of transactions that invalidate
conformance between SHACL shape graph and data graph. Then, the programming
language should be able to type check programs that were written referring to
a defined SHACL shape graph. Type checking should reject programs that could
cause run-time errors, e.g.,\ because they try to access an RDF property that is
not guaranteed to exist without safety precautions. They should also simplify
programs for which queries are guaranteed to return single values rather than
lists, and they should accept programs that do not get stuck when querying
conformant data graphs (with usual exceptions).

To exemplify this, consider three SHACL shapes \texttt{StudentShape},
\texttt{PersonShape} and \texttt{UniversityShape} (see
Fig.~\ref{fig:ex:shacl_intro}).  \texttt{StudentShape} validates all instances
of \texttt{Student}, enforcing that there is at least one \texttt{studiesAt}
relation, that all \texttt{studiesAt} relations point to a node conforming to
the \texttt{UniversityShape} and that all instances of \texttt{Student} are also
instances of \texttt{Person}.  \texttt{PersonShape} validates all instances of
\texttt{Person} and enforces the presence of exactly one \texttt{name} relation.
\texttt{UniversityShape} enforces at least one incoming \texttt{studiesAt}
relation and that all incoming \texttt{studiesAt} relations are from nodes
conforming to the \texttt{StudentShape}.  
\begin{figure}[tb!]
\begin{subfigure}{0.6\textwidth}
\begin{lstlisting}[xleftmargin=2em,numbers=left,basicstyle=\ttfamily]
ex:StudentShape a sh:NodeShape;
  sh:targetClass ex:Student;
  sh:property [
    sh:path ex:studiesAt;
    sh:minCount 1;
    sh:node ex:UniversityShape ];
  sh:class ex:Person.  

ex:PersonShape a sh:NodeShape;
    sh:targetClass ex:Person;
    sh:property [
      sh:path ex:name;
      sh:minCount 1;
      sh:maxCount 1 ].
\end{lstlisting}
\end{subfigure}
\begin{subfigure}{0.4\textwidth}
\begin{lstlisting}[numbers=left,mathescape=true,firstnumber=15]
ex:UniversityShape a 
    sh:NodeShape; 
   sh:property [
    sh:path [
      sh:inversePath;
      ex:studiesAt ];
    sh:minCount 1;
    sh:node 
        ex:StudentShape ].
$~$
$~$
$~$
$~$
$~$
\end{lstlisting}
\end{subfigure}
\caption{SHACL constraints for RDF data graph $G_1$.}
\label{fig:ex:shacl_intro}
\end{figure}
In order for $G_1$ to be valid with respect to the SHACL constraints above,
either the statement that \texttt{alice} is an \texttt{Student} must be removed
or a place of study for \texttt{alice} added. With these changes, the program
above cannot fail anymore.   
A different program (see Lst.~\ref{lst:intro_age}) may query for all instances
of \texttt{Student}. The program may then try to access the \texttt{age}
relation of each query result. However, since it is possible to construct an RDF
graph that is validated by the shapes above, but lacks an \texttt{age} relation
on some instances of \texttt{Student}, the program is unsafe and may crash with
a run-time error. 
\begin{lstlisting}[language=lambda,caption={Program that may produce a run-time
error.},label=lst:intro_age, float]
map (fun x -> x.?X.age) (query { 
    <@\texttt{\textcolor{olive}{SELECT ?X WHERE \{ ?X rdf:type ex:Student.\}}}@> })
\end{lstlisting}
Contrary to that, a similar program that accesses the \texttt{name} relation
instead is guaranteed to never cause run-time errors. 

\paragraph*{Contributions} We propose a type checking procedure based on SHACL
shapes being used as types. We assume that a program queries an---possibly
evolving---RDF data graph that is validated by a SHACL shape graph. Our
contributions are then as follow:
\begin{enumerate}
    \item We define how SHACL shapes can be inferred from queries. As queries
        are the main interaction between programs and RDF data graphs, inferring
        types from data access is a major step in deciding which operations are
        safe. 
    \item 
      We then use a tiny core calculus that captures essential mechanisms to
      define a type system. Due to its simplicity, we use a 
      simply typed $\lambda$-calculus whose basic model of
        computation is extended with queries. We define how SHACL shapes are
        used to verify the program through a type system and show that the
        resulting language is type-safe. That is, a program that passed type
        checking successfully does not yield run-time errors (with the usual
        exception of e.g., accessing the head of an empty list).
\end{enumerate}

\paragraph*{Organization} The paper first recalls basic syntax and semantics for
SPARQL and SHACL in Section~\ref{sec:preliminaries}. Then, the paper describes
how we infer SHACL shapes from queries in
Sections~\ref{sec:shape_inference_revised} and \ref{sec:inference_soundness}
before defining syntax and evaluation rules of the $\lambda$-calculus in
Section~\ref{sec:core_language}. Then, the type system including subtyping is
defined in Section~\ref{sec:type_system} before showing its soundness in
Section~\ref{sec:type_soundness}. Finally, we discuss related work in
Section~\ref{sec:related_work} and conclude in Section~\ref{sec:summary}.

\section{Preliminaries}
\label{sec:preliminaries}


\subsection{SPARQL}
\label{sec:sparql}

RDF graphs are queried via the SPARQL
standard~\cite{sparql}.
We focus on
a core fragment of SPARQL that features conjunctive queries (CQ) and simple path
(P) expressions. We abbreviate this fragment by PCQ. That is, our queries are
conjunctions of property path expressions that use variables only in place of
graph nodes, not in place of path expressions\footnote{As we use plain RDF, we
    do not differentiate between distinguished and existential
variables.}~\cite{sparqlQueryRewriting}. This is also a very widely used subset
of SPARQL queries~\cite{indexingTriples}.    

\paragraph*{Syntax}
\label{sec:sparql_syntax}

We denote the set of graph nodes of an RDF graph $G$ by $N_G$ with $v\in N_G$
denoting a graph node. Furthermore, we assume the existence of a set of
variables $N_V$ with $x$ representing members of this set.  The metavariable $r$
denotes a SPARQL property path expression. A property path expression allows for
defining paths of arbitrary length through an RDF graph.  In our case,
a property path is either a simple iri ($i$), the inverse of a path ($r^{-}$) or
a path that connects subject to object via one or more occurrences of $r$
($r^{+}$). Lastly, we allow for path sequences ($r_1 / r_2$).  A PCQ $q
= (\overline{x}) \leftarrow \mathit{body}$ consists of a head $(\overline{x})$
and a $\mathit{body}$.  We use $\overline{x}$ to denote a sequence of variables
$x_1, \ldots, x_n$. In a head of a PCQ $(\overline{x})$, the sequence
$\overline{x}$ represents the answer variables of the query which are a subset
of all variables occurring in the body of $q$. We use \vars{q} to refer to the
set of all variables occurring in $q$.  Fig.~\ref{fig:sparql_syntax} summarizes
the syntax.  

\begin{figure}[h!t]
\centering
\begin{tabular}{l p{6.0cm}}
    $q$ ::= & $(\overline{x}) \leftarrow \mathit{body}$ \hfill
    ($\mathit{query}$) \\
             & \\
    $\mathit{body}$ ::= & \hfill ($\mathit{query~body}$) \\ 
                         & $\mathit{body} \wedge \mathit{body}$ \hfill
                         (conjunction) \\
                         & | $\mathit{pattern}$ \hfill (pattern) \\
   	& \\
	$\mathit{pattern}$ ::= & \hfill $\mathit{(pattern)}$ \\ 
                            & $x~r~v$ \hfill (subject var pattern) \\
	                        & | $v~r~x$ \hfill (object var pattern) \\
                            & | $x~r~x$ \hfill (subject object var pattern) \\
    & \\
\ifShort{
    $r$ ::= & $i$ | $r^{-}$ | $r / r$ | $r^{+}$ \hfill
        ($\mathit{path~expresssions}$) \\ 
    }
    {
    $r$ ::= & \hfill ($\mathit{property~path~expression}$) \\
            & $i$ \hfill (iri) \\
            & | $r^{-}$ \hfill (inverse path) \\
            & | $r_1 / r_2$ \hfill (path concatenation) \\
            & | $r^{+}$ \hfill (one or more occurrences) \\
    }
\end{tabular}
\caption{Syntax of PCQs.}
\label{fig:sparql_syntax}
\end{figure}

\paragraph*{Semantics}

\ifShort{

For query evaluation, we follow standard semantics. 
Evaluation of a query over a graph $G$ is denoted by $\llbracket \cdot
\rrbracket_G$ and yields a set of mappings $\mu$, mapping variables of the query
onto graph nodes. The full evaluation rules can be found in the extended
technical report of the paper\footnote{Available on arxiv.org.}. 

}
    {

Evaluating a query $q$, follows standard semantics. We use $r(G)$ to denote the
evaluation of a property path expression $r$ on a RDF graph $G$, which consists
of all $(v,v')$ in $G$ such that there is a path from $v$ to $v'$ satisfying
$r$.  Evaluation of $q$ requires the definition of a \emph{mapping} $\mu$.
A mapping $\mu$ is a function $\mu : \vars{q} \rightarrow N_G$ mapping variables
to graph nodes. We use $\Omega$ to denote sets of mappings. The domain
$\mathit{dom}$ of $\mu$ is the subset of $N_V$ where $\mu$ is defined. Two
mappings $\mu_1$ and $\mu_2$ are called \emph{compatible} if for all $x \in
\dom{\mu_1} \cap \dom{\mu_2}$, it is the case that $\mu_1(x) = \mu_2(x)$.
Lastly, to model projection of an query answer $\mu$ onto the answer variables
$\overline{x}$, we use function restriction $\mu_{|\overline{x}}$ to express
that $\mu$ is being restricted to the smaller domain $\overline{x}$. The
evaluation of a query $q$ over a graph $G$, denoted $\llbracket \cdot
\rrbracket_G$ can then be defined as follows:
\begin{description}
    \item $\llbracket x~r~v \rrbracket_G = \{ \mu \mid (\mu(x),v) \in r(G)\}$ 
        \hfill (Q-SVAR)
    \item $\llbracket v~r~x \rrbracket_G = \{ \mu \mid (v,\mu(x)) \in r(G)\}$ 
        \hfill (Q-OVAR)
    \item $\llbracket x_1~r~x_2 \rrbracket_G = \{ \mu \mid (\mu(x_1),\mu(x_2)) 
        \in r(G) \}$ \hfill (Q-VARS)
    \item $\llbracket \mathit{body}_1 \wedge \mathit{body}_2 \rrbracket_G = 
        \llbracket \mathit{body}_1 \rrbracket_G \bowtie 
        \llbracket \mathit{body}_2 \rrbracket_G$ \hfill (Q-CONJ) \\
	    where $\Omega_1 \bowtie \Omega_2 = \{ \mu_1 \cup \mu_2 \mid 
            \mu_1 \in \Omega_1, \mu_2 \in \Omega_2 
            \text{ are compatible mappings} \}$ 
    \item $\llbracket q(\overline{x}) \leftarrow \mathit{body} \rrbracket_G = 
        \{ \mu_{|\overline{x}} \mid \mu \in \llbracket \mathit{body}
        \rrbracket_G \}$ \hfill (Q-PROJ) \\
\end{description}
As an example, consider the following query:  
\[
    q_1 = (x_1,x_2) \leftarrow x_1~\texttt{type}~\texttt{Student} \wedge 
        x_1~\texttt{studiesAt}~x_2
\]
Evaluation of the query against the $G_1$ then looks as follows:
\begin{align*}
    (1) & \llbracket q_1 \rrbracket_{G_1}  
        &=_\text{\tiny{Q-PROJ}}~& 
        \{\mu_{|x_1,x_2} \mid \mu \in (2)\} \\
(2) & \llbracket x_1~\texttt{type}~\texttt{Student} \wedge
    x_1~\texttt{studiesat}~x_2 \rrbracket_{G_1} 
    &=_\text{\tiny{Q-CONJ}}~& (3) \bowtie (4) \\
    &  
    &&=\{\mu_3 = \{(x_1,\texttt{bob}), (x_2,b_1)\}\} \\ 
(3) & \llbracket x_1~\texttt{type}~\texttt{Student}) \rrbracket_{G_1} 
    &=_\text{\tiny{Q-SVAR}}~& 
        \{\mu_1 = \{(x_1,\texttt{alice})\}, \\ 
    & & & ~~\mu_2=\{(x_1,\texttt{bob})\}\} \\     
(4) & \llbracket x_1~\texttt{studiesAt}~x_2) \rrbracket_{G_1} 
    &=_\text{\tiny{Q-VARS}}~& 
        \{\mu_3 = \{(x_1,\texttt{bob}), (x_2,b_1)\}\} \\  
\end{align*}
Evaluation of $x_1~\texttt{type}~\texttt{Student}$ yields two mappings $\mu_1$
and $\mu_2$ that map $x_1$ to \texttt{alice} and \texttt{bob} (3). Evaluation of
$x_1~\texttt{studiesAt}~x_2$ yields a single mapping $\mu_3$ in which $x_1$ is
mapped to \texttt{bob} and $x_2$ is mapped to $b_1$ (4). Joining the mappings
(2) however is only possible for $\mu_2$ and $\mu_3$. $\mu_1$ and $\mu_3$ are
not compatible as they map $x_1$ to different values. Therefore, the query
yields a single result $\mu_3$ in which $x_1$ is mapped to $\texttt{bob}$ and
$x_2$ to $b_1$.

}

\end{document}

\subsection{Shapes Constraint Language (SHACL)}
\label{sec:shacl}

The Shapes Constraint Language (SHACL) is a W3C standard for validating RDF
graphs. In the following, we rely on the definitions presented
by~\cite{SHACL-Semantics}. SHACL groups constraints in so-called \emph{shapes}.  A shape is
referred to by a name, it has a set of constraints and defines its \emph{target
nodes}.  Target nodes are those nodes of the graph that are expected to fulfill
the constraints of the shape. As exemplified by \texttt{StudentShape} and
\texttt{UniversityShape} (see Fig.~\ref{fig:ex:shacl_intro}), constraints may
reference other shapes. 

\paragraph*{Constraint Syntax}
\label{sec:constraint_syntax}

We start by defining constraints. We follow~\cite{SHACL-Semantics}, who use
a logical abstraction of the concrete SHACL language. Fragments of first order
logic are used to simulate node shapes whereas so called property shapes are
completely abstracted away.  Constraints that are used in shapes are defined by
the following grammar: 
\begin{equation}
    \phi ::= \top \mid s \mid v \mid \phi_1 \wedge \phi_2 \mid \neg \phi \mid
        \geq_n r.\phi  
\end{equation}
where $s$ is a shape name (indicating a reference to another shape), $v$ is
a constant (or rather a graph node), $r$ is a property path and
$n\in\mathbb{N}^{+}$.  Additional syntactic constructs may be derived from this
basic grammar, including $\leq_n r.\phi$ for $\neg(\geq_{n+1} r.\phi)$, $=_n
r.\phi$ for $(\leq_n r.\phi) \wedge (\geq_n r.\phi)$ and $\phi_1 \lor \phi_2$
for $\neg (\neg \phi_1 \land \neg \phi_2)$. We sometimes use $\phi_s$ to denote
the constraint belonging to a specific shape $s$. To improve readability, we
sometimes add parenthesis to constraints although they are not explicitly
mentioned in the grammar. 
%

\paragraph*{Constraint Evaluation}
\label{sec:constraint_evaluation}

\ifShort{
Evaluation of constraints is rather straightforward with the exception of
reference cycles. Evaluation is therefore grounded using assignments $\sigma$
which map graph nodes to shape names~\cite{SHACL-Semantics}. We rely on total
assignments instead of partial assignments for simplicity. 
\begin{definition}[Total Assignment] Let $G$ be an RDF data graph with its set
    of nodes $N_G$ and let $N_S$ a set of shape names. Then $\sigma$ is a total
    function $\sigma : N_G \rightarrow 2^{N_S}$ mapping graph nodes $v$ to
    subsets of $N_S$. If $s \in \sigma(v)$, then $v$ is assigned to the shape $s$. 
\end{definition}

Evaluation of a constraint $\phi$ for a node $v$ of a graph $G$ using an
assignment $\sigma$ is denoted $\llbracket \phi \rrbracket^{v,G,\sigma}$ and
yields either true or false. The extended version contains the complete
definition.  
}
    {

Evaluation of constraints is rather straightforward with the exception of
reference cycles. To highlight the issues with reference cycles, consider
a shape name $s_\mathit{local}$ with its constraint $\leq_0 \texttt{knows}.\neg
s_\mathit{local}$. In order to fulfill constraints of the ``LocalShape'', one
must only know other locals. Furthermore, consider a graph consisting of
a single vertex $b_1$ who knows itself (see Fig.~\ref{fig:ex:local_shape}).  
\begin{figure}[h!]
\begin{subfigure}{0.45\textwidth}
    \centering
    \[ \phi_{\mathit{local}} = (\leq_0 \texttt{knows}.\neg s_{\mathit{local}}) \]
\end{subfigure}
\begin{subfigure}{0.45\textwidth}
    \centering
    \begin{tikzpicture}[auto,scale=1.0,transform shape]

        \begin{scope}[every node/.style={font=\ttfamily\footnotesize}]
            \node (v) at (0,0) {$b_1$};
        \end{scope}

        \begin{scope}[every edge/.style={draw=black},
            every node/.style={font=\ttfamily\footnotesize}]
            \draw [->] (v) edge[loop right] node{knows} (v);
        \end{scope}
    \end{tikzpicture}    
\end{subfigure}
\caption{Illustration of a problematic, recursive case.}
\label{fig:ex:local_shape}
\end{figure}
Intuitively, there are two possible solutions. If $b_1$ is assumed to conform to
$s_\mathit{local}$, then the constraint is fulfilled and the assumption is
justified. Likewise, if $b_1$ is assumed to not conform to the
$s_\mathit{local}$ shape, then the constraint is violated and it is correct to
say that $b_1$ does not conform to $s_\mathit{local}$. 

As introduced by~\cite{SHACL-Semantics}, we ground evaluation using an
assignment $\sigma$ to resolve this issue. An assignment $\sigma$ assigns graph
nodes $v$ to shape names $s$. Evaluation of constraints takes an assignment as
a parameter and evaluates the constraints with respect to the given assignment.
The case above is therefore represented through two different assignments---one
in which $s_\mathit{local} \in \sigma_1(b_1)$ and a different one where
$s_\mathit{local} \not \in \sigma_2(b_1)$. We require total assignments that map
all graph nodes to the set of all shapes that the node supposedly conforms to.  
We use $N_S$ to denote the set of SHACL shape names: 

\begin{definition}[Total Assignment] Let $G$ be an RDF data graph with its set
    of nodes $N_G$ and let $N_S$ a set of shape names. Then $\sigma$ is a total
    function $\sigma : N_G \rightarrow 2^{N_S}$ mapping graph nodes $v$ to
    subsets of $N_S$. If $s \in \sigma(v)$, then $v$ is assigned to the shape $s$. 
    For all $s \not\in \sigma(v)$, the node $v$ is not assigned to the shape $s$.  
\end{definition}

Evaluating whether a graph node $v$ in a given RDF graph $G$ satisfies
a constraint $\phi$, written $\llbracket \phi \rrbracket^{v, G, \sigma}$ can
then be defines as shown in Fig.~\ref{fig:constraint_evaluation}: 
\begin{figure}[h!t]
\centering
\begin{boxedminipage}{1\textwidth}
\begin{align*}
    \llbracket \top \rrbracket^{v, G, \sigma} &= \mathit{true} \\
    \llbracket \neg \phi \rrbracket^{v, G, \sigma} &= 
        \text{not} \llbracket \phi \rrbracket^{v, G, \sigma} \\
    \llbracket \phi_1 \wedge \phi_2 \rrbracket^{v, G, \sigma} &=
    \llbracket \phi_1 \rrbracket^{v, G, \sigma} \text{ and }
            \llbracket \phi_2 \rrbracket^{v, G, \sigma} \\
    \llbracket v' \rrbracket^{v, G, \sigma} &= \begin{cases}
        \mathit{true}, \text{ if $v = v'$} \\
        \mathit{false}, \text{ otherwise} \\
    \end{cases} \\
    \llbracket s \rrbracket^{v, G, \sigma} &= \begin{cases} 
        \mathit{true}, \text{ if } s \in \sigma(v) \\
        \mathit{false}, \text{ if } s \not\in \sigma(v) \\
    \end{cases} \\
    \llbracket \geq_n r.\phi \rrbracket^{v, G, \sigma} &=
    \begin{cases}
        \mathit{true}, \text{ if } |\{v'|(v,v')\in r(G) \text{ and } 
            \llbracket \phi \rrbracket^{v', G, \sigma} = \mathit{true}\}| 
            \geq n \\
        \mathit{false}, \text{ otherwise} \\
    \end{cases} 
\end{align*}

\end{boxedminipage}
\caption{Evaluation rules of constraints.}
\label{fig:constraint_evaluation}
\end{figure}

To illustrate constraint evaluation, consider the representation of
``UniversityShape'', $\phi_\mathit{University} = (\geq_1
\texttt{studiesAt}^{-}.s_\mathit{Student} \wedge \geq_1
\texttt{locatedIn}.\top)$ again. For $G_1$, an assignment $\sigma_1$ may for
example assign $s_\mathit{Student}$ to the node \texttt{bob}
($s_\mathit{Student} \in \sigma_1(\texttt{bob})$). The evaluation of
$\phi_\mathit{University}$ for the node $b_1$ using $\sigma_1$ then looks as
follows:
\begin{align*}
(1) & \llbracket \phi_\mathit{University} \rrbracket^{b_1,G_1,\sigma_1} 
    &=~&
    (2) \text{ and } (4) \\
(2) & \llbracket \geq_1 \texttt{studiesAt}^{-}.s_\mathit{Student}
        \rrbracket^{b_1,G_1,\sigma_1}
    &=~& |\{ \texttt{bob} \}| \geq 1~\text{because} \\
    &
    &&~\texttt{studiesAt}^{-}(G_1) = \{(b_1,\texttt{bob})\}~\text{and}~(3)\\
(3) & \llbracket s_\mathit{Student} \rrbracket^{\texttt{bob},G_1,\sigma_1} 
    &=~& s_\mathit{Student} \in \sigma_1(\texttt{bob}) \\
\end{align*}

To evaluate the constraint, both $\geq_1
\texttt{studiesAt}^{-}.s_\mathit{Student}$ and $=_1 \texttt{locatedIn}.\top$
must evaluate to true (1). For the first part, the set of all nodes with
\texttt{studiesAt} relations pointing to $b_1$ ($(b_1,v') \in
\texttt{studiesAt}^{-}(G_1)$) is constructed. The set consists solely of
\texttt{bob}. Then, it is checked whether \texttt{bob} is assigned to the
$s_\mathit{Student}$ shape (3). Since \texttt{bob} is, he is kept in the set and
the constraint evaluates to true. 

}

\paragraph*{Shapes and Validation}
\label{sec:shape_validaton}

\ifShort{

A shape is modelled as a triple $(s, \phi, q)$ consisting of a shape name $s$,
a constraint $\phi$ and a query for target nodes $q$ which is either an
empty set or a monadic query that has exactly one answer variable to describe
all intended \emph{target nodes}. Target nodes denote those nodes which are
expected to fulfill the constraint associated with the shape. In a slight abuse of notation, we write $v
\in \llbracket q \rrbracket_G$ to indicate that a node $v$ is a target node
for $s$ in the graph $G$. If $S$ is a set of shapes, we assume that for each
$(s, \phi, q) \in S$, if shape name $s'$ appears in $\phi$, then there
also exists a $(s',\phi',q') \in S$. 
To illustrate this, consider our running example again (see
Fig.~\ref{fig:ex:shacl_intro})\footnote{We simplified target
queries in the example---in reality, the target queries should query for Student or any of its
subclasses. We simplified this as we do not use any RDFS subclass relations in
our examples.}.  The set $S_1$
containing all three shapes looks as follows:  
{\small
\begin{align*}
    S_1 =\{ & (s_\mathit{Student},
                \geq_1 \texttt{studiesAt}.\top \wedge \leq_0
                \texttt{studiesAt}.\neg s_\mathit{University} \wedge \geq_1
                \texttt{type}.\texttt{Person},
                \\
            & \hspace{0.5cm} (x_1) \leftarrow x_1~\texttt{type}~\texttt{Student}), \\
            & (s_\mathit{Person},
                =_1 \texttt{name}.\top,
                (x_1) \leftarrow x_1~\texttt{type}~\texttt{Person}), \\
            & (s_\mathit{University},
                \geq_1 \texttt{studiesAt}^{-}.\top \wedge \leq_0
                \texttt{studiesAt}^{-}.s_\mathit{Student},
                \emptyset) \} 
\end{align*}
}%
Intuitively, only certain assignments are of interest. 
Such an assignment is called a \emph{faithful assignment}.   

\begin{definition}[Faithful assignment]  
An assignment $\sigma$ for a graph $G$ and a set of shapes $S$ is faithful, iff
for each $(s,\phi,q) \in S$ and for each graph node $v \in N_G$, it holds that: 
\begin{itemize}
    \item if $v \in \llbracket q \rrbracket_G$, then $s \in \sigma(v)$. 
    \item if $s\in\sigma(v)$, iff $\llbracket \phi 
        \rrbracket^{v, G, \sigma} = \mathit{true}$.
\end{itemize}
\end{definition}

Validating an RDF graph means finding a faithful assignment. The graph is said
to \emph{conform} to the set of shapes. 
\begin{definition}[Conformance] A graph $G$ conforms to a set of shapes $S$ iff
there is a faithful assignment $\sigma$ for $G$ and $S$. We write $\sigma^{G,S}$
to denote that $\sigma$ is a faithful assignment for $G$ and $S$.
\end{definition}
Validating an RDF graph means finding a faithful assignment. In case of graph $G_1$ (see
Fig.~\ref{fig:ex:g1}) and the set of shapes $S_1$, it is impossible to validate
the graph. \texttt{alice} would need to be assigned to $s_\mathit{Student}$, but
has no \texttt{studiesAt} relation. However, if the statement
(\texttt{alice},\texttt{type},\texttt{Student}) is removed, then the graph is
valid since a faithful assignment may assign $s_\mathit{Person}$ to
\texttt{alice} and \texttt{bob}, $s_\mathit{Student}$ solely to \texttt{bob} and
$s_\mathit{University}$ to $b_1$.

}
    {

A shapes is modelled as a triple $(s, \phi, q)$ consisting of a shape name $s$,
a constraint $\phi_s$ and a query for target nodes $q_s$ which is either an
empty set or a monadic query that has exactly one answer variable to describe
all intended \emph{target nodes}. Target nodes denote those nodes which should
be evaluated against the constraint and which are expected to fulfill the
constraint associated with the shape. In a slight abuse of notation, we write $v
\in \llbracket q_s \rrbracket_G$ to indicate that a node $v$ is a target node
for $s$ in the graph $G$. If $S$ is a set of shapes, we assume that for each
$(s, \phi_s, q_s) \in S$, if shape name $s'$ appears in $\phi_s$, then there
also exists a $(s',\phi_{s'},q_{s'}) \in S$. 

To illustrate this, consider our running example again (see Fig.~\ref{fig:ex:shacl}). 
The set $S_1$ containing all three shapes looks as follows:  
{\small
\begin{align*}
    S_1 =\{ & (s_\mathit{Student}, 
                \geq_1 \texttt{studiesAt}.s_\mathit{University}\, \wedge \geq_1
                \texttt{type}.\texttt{Person} , 
                (x_1) \leftarrow x_1~\texttt{type}~\texttt{Student}), \\
            & (s_\mathit{Person},
                =_1 \texttt{name}.\top,
                (x_1) \leftarrow x_1~\texttt{type}~\texttt{Person}), \\
            & (s_\mathit{University}, 
                \geq_1 \texttt{studiesAt}^{-}.s_\mathit{Student},
                \emptyset) \} 
\end{align*}
}%

Intuitively, when validating an RDF graph with the set of shapes $S_1$, only
certain assignments are of interest. For one, due to the target nodes of the
shape, any assignment that could validate the graph should assign all instances
of \texttt{Student} to the shape $s_\mathit{Student}$ and all instances of
\texttt{Person} to $s_\mathit{Person}$. Second, if an assignment assigns
a shape to a graph node, the constraint of the shape should evaluate to true.
Such an assignment is called a \emph{faithful assignment}.   

\begin{definition}[Faithful assignment]  
An assignment $\sigma$ for a graph $G$ and a set of shapes $S$ is faithful, iff
for each $(s,\phi_s,q_s) \in S$ and for each graph node $v \in N_G$, it holds that: 
\begin{itemize}
    \item if $v \in \llbracket q_s \rrbracket_G$, then $s \in \sigma(v)$. 
    \item if $s\in\sigma(v)$, then $\llbracket \phi_s 
        \rrbracket^{v, G, \sigma} = \mathit{true}$.
    \item if $s \not\in\sigma(v)$, then $\llbracket \phi_s 
        \rrbracket^{v, G, \sigma} = \mathit{false}$.  
\end{itemize}
\end{definition}

Lastly, if a faithful assignment can be found for an RDF graph, it is possible
to validate the graph---that is, the graph fulfills all constraints given by the
set of shapes. The graph is said to \emph{conform} to the set of shapes.
\begin{definition}[Conformance] A graph $G$ conforms to a set of shapes $S$ iff
there is a faithful assignment $\sigma$ for $G$ and $S$. We write $\sigma^{G,S}$
to denote that $\sigma$ is a faithful assignment for $G$ and $S$.
\end{definition}

Validating an RDF graph means finding a faithful assignment. It is akin to
checking for satisfiability in logics. Finding a faithful assignment may not
necessary be possible. In case of graph $G_1$ (see Fig.~\ref{fig:ex:g1}) and the
set of shapes $S_1$, it is impossible to validate the graph. \texttt{alice}
would need to be assigned to $s_\mathit{Student}$, but has no \texttt{studiesAt}
relation---therefore the constraint does not evaluate to true. However, if the
statement (\texttt{alice},\texttt{type},\texttt{Student}) is removed, then the
graph is valid since a faithful assignment may assign $s_\mathit{Person}$ to
\texttt{alice} and \texttt{bob}, $s_\mathit{Student}$ solely to \texttt{bob} and
$s_\mathit{University}$ to $b_1$.

Due to negation, some reference cycles cannot be satisfied. As an example,
consider a set of shapes $S$ for which $(s_\mathit{unsatisfiable}, \neg
s_\mathit{unsatisfiable}, \emptyset) \in S$. To satisfy the constraint and
conform to $s_\mathit{unsatisfiable}$, one would need to not conform to the shape.
$s_\mathit{unsatisfiable}$ makes it impossible to conform to the set of shapes
$S$. To avoid such cases, we only consider sets of shapes in which constraints
can be stratified to ensure that negation and reference cycles are used in
a sensible manner.
\begin{definition}[Stratification] A set of shapes $S$ with  $s_1,s_2 \in S$ is
stratified if there is a total function $\mathit{str}: S \rightarrow
\mathbb{N}$ such that: 
\begin{itemize} 
    \item If $s_1$ appears in $\phi_{s_2}$, then $\mathit{str}(s_1) \leq
        \mathit{str}(s_2)$.  
    \item If $s_1$ appears in $\phi_{s_2}$ in the scope of a negation, then
        $\mathit{str}(s_1) < \mathit{str}(s_2)$.  
\end{itemize} 

\end{definition}

}

\end{document}


\section{Shape Inference for Queries}
\label{sec:shape_inference_revised}

In this section, we describe how to infer shapes from PCQs for all variables in
a given query. Given a query $q$ with $x\in vars(q)$, let $s^q_x$ be the
globally unique shape name for variable x in query q. Then we assign the shape
$(s_x^q, \phi, q_x)$. We discard sub- or superscripts if they are evident in
context.

Our typing relation ``:'' for a PCQ $q$ constructs a set of shapes $S_q$ in the
following manner: For every subject var pattern $x~r~v$ in the body of $q$
(object var pattern $v~r~x$ respectively), we assign the constraint $\geq_1
r.v$ ($\geq_1 r^{-}.v$). As target nodes, we use the original query but
projected on the particular variable.
In case of variables on both subject and object ($x_1~r~x_2$), we infer two
shapes $s^q_{x_1}$ and $s^q_{x_2}$.  We use shape references to express the
dependencies and infer the constraints $\geq_1 r.s^q_{x_2}$ and $\geq_1
r^{-}.s^q_{x_1}$. In case of a conjunction ($\mathit{body}_1 \wedge
\mathit{body_2}$), we infer the sets of constraints for each query body
individually and then combine the results using the operator $\boxbowtie$. 
The relation $\boxbowtie$ takes two sets of shapes $S_{q_1}$ and $S_{q_2}$
combines them into a unique set performing a full outer join on the shape names:
{\small
\begin{align*}
    S_{q_1} \boxbowtie S_{q_2} =& 
    \{ (s^q_{x_i},\phi_i \wedge \phi_j, 
        (x_i) \leftarrow \mathit{body}_i \wedge \mathit{body}_j) |  
        (s^q_{x_i}, \phi_i, (x_i) \leftarrow \mathit{body}_i) \in S_{q_1} \\ 
    & \wedge (s^q_{x_i}, \phi_j, (x_i) \leftarrow
                            \mathit{body}_j) \in S_{q_2} \} 
        \; \cup \\
    & \{(s^q_{x_i}, \phi_i, q_i) | 
        (s^q_{x_i}, \phi_i, q_i) \in S_{q_1} \wedge
        \neg\exists (s^q_{x_i}, \phi_j, q_j) \in S_{q_2}\} \; \cup \\
    & \{(s^q_{x_j}, \phi_j, q_j) | 
        \neg \exists(s^q_{x_j},\phi_i, q_i) \in S_{q_1} \wedge 
        (s^q_{x_j},\phi_j, q_j) \in S_{q_2} \}
\end{align*}
}%
Fig.~\ref{fig:rules:shape_inference} contains the complete set of rules for
inferring sets of shapes from PCQs. 

\begin{figure}[h!t]
\begin{boxedminipage}{1\textwidth}
\begin{mathpar}
        
    \infer{}{x~r~v : \{(s^q_x,
            \geq_1 r.v,
            (x) \leftarrow x~r~v)\}}~\text{(R-SUB-VAR)}
    
    \infer{}{v~r~x : \{(s^q_x,
            \geq_1 r^{-}.v,
            (x) \leftarrow v~r~x)\}}~\text{(R-OBJ-VAR)}

    \infer{}{x_1~r~x_2 : \{
            (s^q_{x_1},
             \geq_1 r.s^q_{x_2},
             (x_1) \leftarrow x_1~r~x_2),  
            (s^q_{x_2}, 
             \geq_1 r^{-}.s^q_{x_1},
             (x_2) \leftarrow x_1~r~x_2) \}}
             ~\hspace{-0.0cm}\text{(R-VARS)}

    \infer{\mathit{body}_1 : S_{q_1} \\ \mathit{body}_2 : S_{q_2}}
        {\mathit{body}_1 \wedge \mathit{body}_2 : 
        S_{q_1} \boxbowtie S_{q_2}}~\text{(R-CONJ)}

    \infer{\mathit{body} : S_q}
    {(\overline{x}) \leftarrow \mathit{body} : S_q}~\text{(R-PROJ)}
\end{mathpar}
\end{boxedminipage}
\caption{Inference rules for inferring a set of shapes from the body of query
$q$.}
\label{fig:rules:shape_inference}
\end{figure}
As an example, consider the query $q = (x_1,x_2) \leftarrow
x_1~\texttt{type}~\texttt{Student}~\wedge~x_1~\texttt{studiesAt}~x_2$ as used 
before. 
Then shape inference on the body assigns the following set of shapes:
{\small
\begin{align*}
(1) &~ x_1~\texttt{type}~\texttt{Student}~\wedge~x_1~\texttt{studiesAt}~x_2 : (2) \boxbowtie (3)   \\
    &~                = \{(s^q_{x_1},\geq_1 \texttt{type.Student} \wedge 
        \texttt{studiesAt}.s^q_{x_2}, (x_1) \leftarrow x_1~\texttt{type}~\texttt{Student}~\wedge~x_1~\texttt{studiesAt}~x_2), \\
    &\hspace{0.75cm} (s^q_{x_2},\geq_1 \texttt{studiesAt}^{-}.s^q_{x_1},
        (x_2)\leftarrow x_1~\texttt{type}~\texttt{Student}~\wedge~x_1~\texttt{studiesAt}~x_2) \} \\
(2) &~ x_1~\texttt{type}~\texttt{Student} : \{(s^q_{x_1},\geq_1
        \texttt{type.Student},(x_1) \leftarrow x_1~\texttt{type}~\texttt{Student} )\} \\
(3) &~ x_1~\texttt{studiesAt}~x_2 : \{(s^q_{x_1}, \geq_1
    \texttt{studiesAt}.s^q_{x_2},(x_1) \leftarrow x_1~\texttt{studiesAt}~x_2),\,~~~ \\
    &\hspace{2.75cm}(s^q_{x_2}, \geq_1 \texttt{studiesAt}^{-}.s^q_{x_1}, (x_2) \leftarrow x_1~\texttt{studiesAt}~x_2)\} 
\end{align*}
}%

\section{Soundness of Shape Inference for Queries}
\label{sec:inference_soundness}

Shape inference for queries is sound if the shape constraints inferred for each
variable evaluate to true for all possible mappings of the variable.
\begin{definition}[Soundness of shape inference]
Given an RDF graph $G$, a PCQ $q$ with its variables $x_i \in \vars{q}$ and the
set of inferred shapes $S_q = \{ (s^q_{x_i}, \phi_{x_i}, q_{s_{x_i}})^{x_i \in
\vars{q}} \}$, a shape constraint is sound if there exists a faithful assignment
$\sigma^{G,S_q}$ such that
\[ 
\forall x_i \in \vars{q}: 
\forall \mu \in \llbracket q \rrbracket_G: 
\llbracket \phi_{x_i} \rrbracket^{\mu(x_i), G, \sigma^{G,S_q}} = \mathit{true} 
\]  
\end{definition}
We show that the faithful assignment $\sigma^{G,S_q}$ can be constructed by
assigning all shape names solely based on target nodes. 

\begin{theorem}\label{th:soundness_assignment}
For any graph $G$, a PCQ $q$ and the set of shapes $S_q$ inferred from
$q$, assignment $\sigma^{G,S_q}$ is constructed such that for each shape
$(s,\phi_s,q_s) \in S_q$ and for each graph node $v \in N_G$:
\begin{enumerate}
    \item If $v \in \llbracket q_s \rrbracket_G$, then $s \in \sigma^{G,S_q}(v)$, 
    \item If $v \not\in \llbracket q_s \rrbracket_G$, $s \not\in \sigma^{G,S_q}(v)$. 
\end{enumerate}
Such an assignment $\sigma^{G,S_q}$ is faithful. 
\end{theorem}

\ifShort{

\begin{proof}[Proof (Sketch)]
Intuitively, a node $v$ is part of the query result due to the presence of some
relations for the node. The assigned constraints require the presence of the
exact same relations to evaluate to true. A induction over the query evaluation
rules can therefore show that 1) all nodes that are in the query result fulfill
the constraint whereas 2) a node not being in the query result would also
violate the constraint. 
\end{proof}

}
    {

\begin{proof}
An assignment is faithful if three conditions are met. First, for all
$(s,\phi_s,q_s) \in S_q$ and for all $v \in \llbracket q_s \rrbracket_G$, it
must be that $s \in \sigma^{G,S_q}(v)$. This is fulfilled through the construction of
$\sigma^{G,S_q}$.  Furthermore, it must be true that for all $v \in N_G$:
\begin{enumerate}
    \item if $s \in \sigma^{G,S_q}(v)$, then $\llbracket \phi_s \rrbracket^{v,
        G, \sigma^{G,S_q}} = \mathit{true}$.
    \item if $s \not\in \sigma^{G,S_q}(v)$, then $\llbracket \phi_s \rrbracket^{v,
        G, \sigma^{G,S_q}} = \mathit{false}$. 
\end{enumerate}
We show this by induction on the evaluation of $\llbracket q = (\overline{x})
\leftarrow \mathit{body} \rrbracket_G$.
\begin{description}
    \item[(Q-SVAR)] For the query $\mathit{body} = x~r~v'$, the inferred set of
        shapes $S_q$ is $\{ (s^q_x,\geq_1 r.v', (x) \leftarrow x~r~v')) \}$.
        Evaluation of the query returns $v$ for which $(v,v') \in r(G)$.  
        \begin{enumerate}
            \item The constraint requires all $v$ assigned to shape $s^q_x$ to
                have at least one successor via the relation $r$ pointing to
                $v'$. This is true for all $v$ since they would not be in the
                query result otherwise. Therefore, $s^q_x \in \sigma^{G,S_q}(v)$ as
                required by the construction of $\sigma^{G,S_q}$, does not violate
                faithfulness. 
            \item Any node $v'' \in N_G$ for which $s^q_x \not\in
                \sigma^{G,S_q}(v'')$ must violate the constraint. By design of
                $\sigma^{G,S_q}$, any node $s^q_x \not\in \sigma^{G,S_q}(v'')$
                cannot be part of the query result. This means that they cannot
                have a successor via the relation $r$ pointing to $v'$.
                Therefore, those nodes violate the constraint and
                $\sigma^{G,S_q}$ is faithful.  
        \end{enumerate}
        
    \item[(Q-OVAR)] For the query $\mathit{body} = v~r~x$, the inferred set of
        shapes $S_q$ is $\{(s^q_x, \geq_1 r^{-}.v, (x) \leftarrow v~r~x)\}$.
        This case is similar to case (Q-SVAR).

    \item[(Q-VARS)] For the query $\mathit{body} = x_1~r~x_2$, the inferred set
        of shapes $S_q$ is $\{(s^q_{x_1},\geq_1 r.s^q_{x_2}, (x_1) \leftarrow
        x_1~r~x_2)), (s^q_{x_2}, \geq_1 r^{-}.s^q_{x_1}, (x_2)\leftarrow x_1~r~x_2)
        \}$. Evaluation of the query returns all $(v,v') \in r(G)$
        whereas construction of $\sigma^{G,S_q}$ assigns all $v$ to shape $s^q_{x_1}$ and
        all $v'$ to shape $s^q_{x_2}$.
        \begin{enumerate} 
            \item The constraint requires all $v$ to have at least one successor
                $v'$ via the relation $r$ that is assigned to the shape
                $s^q_{x_2}$.  This is fulfilled through the construction of
                $\sigma^{G,S_q}$. Likewise, all $v'$ require a predecessor via
                $r$ that is assigned to $s^q_{x_1}$.  Again, this must be true
                through the construction of $\sigma^{G,S_q}$.  Therefore, the
                constraints evaluates to true for all $v$ and $v'$ respectively
                and the assignment $\sigma^{G,S_q}$ is still faithful. 

            \item Any node $v'' \in N_G$ for which neither $s^{q}_{x_1}
                \not\in \sigma^{G,S_q}(v'')$ nor $s^{q}_{x_2} \not\in
                \sigma^{G,S_q}(v'')$ cannot have a successor or predecessor via
                the relation $r$ as they would otherwise be part of the query
                result. Both constraints would therefore evaluate to false and
                $\sigma^{G,S_q}$ is still faithful.         
        \end{enumerate}

    \item[(Q-CONJ)] For the query $\mathit{body} = \mathit{body}_1 \wedge
        \mathit{body}_2$, both $\mathit{body}_1$ and $\mathit{body}_2$ infer
        their own set of shapes $S_{q_1}$ and $S_{q_2}$ which are combined into
        $S_q = S_{q_1} \boxbowtie S_{q_2}$. 
        By induction hypothesis, $\sigma^{G,S_q}$ is faithful for $G$ and $S_{q_1}$
        and $S_{q_2}$ individually. Evaluation of the query returns
        $\mathit{body}_1 \bowtie \mathit{body}_2$ evaluates each part
        individually and, for all query results $\mu_1$ and $\mu_2$, takes the
        union in case they are compatible. $\mu_1$ and $\mu_2$ are compatible
        if, for all variables $x \in \dom{\mu_1} \cap \dom{\mu_2}$, it holds
        that $\mu_1(x) = \mu_2(x)$. Therefore, for each variable $x_i$, there are two
        cases to consider:
        \begin{description}
            \item[$x_i$ occuring in both bodies:] $\boxbowtie$ takes the
                conjunction of the constraints for $x_i$ in $S_{q_1}$ and
                $S_{q_2}$. 
                \begin{enumerate} 
                    \item By induction hypothesis, both $\phi_{i_1}$
                    from $(s^q_{x_i},\phi_{i_1},q_{i_1}) \in S_{q_1}$ and
                    $\phi_{i_2}$ from $(s^q_{x_i},\phi_{i_2},q_{i_2}) \in S_{q_2}$
                    evaluate to true for all possible mappings of $x_i$. As
                    $\boxbowtie$ constructs $\phi_{i_1} \wedge \phi_{i_2}$ and no 
                    negation is used in either constraint, the resulting
                    constraint must also evaluate to true. 
                    \item As no negation occurs in constraints of $S_{q_1}$ and
                        $S_{q_2}$, it is impossible for any nodes previously
                        violating any constraints to fulfill the conjunction of
                        the constraints. 
                \end{enumerate}
            \item[$x_i$ only occuring in one body:] The constraint for the
                variable is not modified by $\boxbowtie$. The assignment is
                therefore still faithful. 
        \end{description}
        
    \item[(Q-PROJ)] $q = (\overline{x}) \leftarrow \mathit{body}$, $\mathit{body} : 
        S_q$, $q : S_q$. 
        Immediate since the inferred set of shapes is not modified. 
\end{description}
\end{proof}

}

The faithful assignment $\sigma^{G,S_q}$ constructed in the manner as explained
above is unique. This is expected as shape inference does not use negation.

\begin{proposition}
The assignment $\sigma^{G,S_q}$ constructed as described above is unique.  
\end{proposition}

\begin{proof}
Assume that a different faithful assignment $\sigma'^{G,S_q}$ exists. There must be at
least one node $v$ for which $\sigma^{G,S_q}(v) \neq \sigma'^{G,S_q}(v)$. 
\begin{enumerate}
    \item It is impossible that there is an $s$ such that $s \in
        \sigma^{G,S_q}(v)$ and $s \not\in \sigma'^{G,S_q}(v)$. $\sigma$
        assigns shapes based on target nodes, $v$ must be a target node for $s$
        and $\sigma'$ is not faithful.

    \item It cannot be that $s \not\in \sigma^{G,S_q}(v)$ and $s \in
        \sigma'^{G,S_q}(v)$. $v$ must fulfill the constraint $\phi_s$ of shape $s$,
        otherwise $\sigma'$ would not be faithful. If that is the case, then
        $\sigma$ is not faithful. This contradicts Theorem~\ref{th:soundness_assignment}. 
\end{enumerate}
\vspace{-0.6cm}
\end{proof}

\ifShort{

Given a faithful assignment $\sigma^{G,S}$ for a set of shapes $S$ and
assignment $\sigma^{G,S_q}$ for an inferred set of shapes, the two assignments can be
combined by simply taking the union $\sigma^{G,S}(v) \cup \sigma^{G,S_q}(v)$ for
each graph node $v \in N_G$. While not true for two arbitrary assignments, it is
true in this case because shape names of $S$ and $S_q$ are disjoint.

}
    {

Given a faithful assignment $\sigma^{G,S}$ for a set of shapes $S$, the
assignment $\sigma^{G,S_q}$ for a inferred set of shapes, the two assignments can be
combined through an operator $\Cup$ which, for each graph node $v$, takes the
union of $\sigma^{G,S}(v) \cup \sigma^{G,S_q}(v)$.

This is not true for arbitrary faithful assignments. As a counter example,
consider a set of shapes consisting of $s_\mathit{local}$, who may only know
other locals and $s_\mathit{semilocal}$ who must know at least one node who is
not a local. Given a data graph consisting of one node $b_1$ which knows itself
(see Fig.~\ref{fig:ex:multiple_assignments}), two faithful assignments
$\sigma$ and $\sigma'$ exist.
\begin{figure}[h!]
\begin{subfigure}{0.45\textwidth}
\begin{align*}
    S = \{ & (s_\mathit{local},\leq_0 \texttt{knows}.\neg s_\mathit{local},
        \emptyset), \\ 
           & (s_\mathit{semilocal}, \geq_1 \texttt{knows}.\neg s_\mathit{local},
       \emptyset)\}
\end{align*}
\end{subfigure}
\begin{subfigure}{0.45\textwidth}
    \centering
    \begin{tikzpicture}[auto,scale=1.0,transform shape]

        \begin{scope}[every node/.style={font=\sffamily\footnotesize}]
            \node (v) at (0,0) {$b_1$};
        \end{scope}

        \begin{scope}[every edge/.style={draw=black}]
            \draw [->] (v) edge[loop right] node{knows} (v);
        \end{scope}
    \end{tikzpicture}    
\end{subfigure}
\caption{Basic example for multiple faithful assignments.}
\label{fig:ex:multiple_assignments}
\end{figure}
In assignment $\sigma$, the node $b_1$ is assigned to the shape
$s_\mathit{local}$ but not $s_\mathit{semilocal}$ ($\sigma(b_1)
= \{s_\mathit{local}\}$). Likewise, in assignment $\sigma'$, $b_1$ is only
assigned to $s_\mathit{semilocal}$ but not $s_\mathit{local}$ ($\sigma'(b_1)
= \{ s_\mathit{semilocal} \}$). Individually, both assignments are faithful, but
combining them ($\sigma \Cup \sigma'$) does not yield a faithful
assignment as neither constraint evaluates to true. 

However, in case of $\sigma^{G,S_q}$ for $S_q$, combining it with an other faithful
assignment $\sigma^{G,S}$ for a set of shapes $S$ will yield a faithful assignment
again. This is because shape names of $\sigma^{G,S_q}$ are unique. $S$ cannot contain
a shape $(s,\phi_s,q_s)$ for which $\phi_s$ mentions a shape name $s^q_x$ such
that $(s^q_x,\phi^q_x,q^q_{s_x}) \in S_q$. Combining assignments therefore has no
effect on constraint evaluation.  

\begin{proposition}
The assignment $\sigma^{G,S_q}$ can be combined with any other assignment
$\sigma^{G,S}$
through a operator $\Cup$ that, for each graph node $v$, takes the union of
$\sigma^{G,S_q}$ and $\sigma^{G,S}$: 
\[\forall v\in G: (\sigma^{G,S_q} \Cup \sigma^{G,S})(v) = \sigma^{G,S_q}(v) \cup
\sigma^{G,S}(v)\]  
\end{proposition}

\begin{proof} 
Shape names in $\sigma_{G,S_q}q$ are completely disjunct from shape names in
$\sigma_{G,S}$ and therefore have no effect on the evaluation of constraints. 
\end{proof}

}

\end{document}

\section{Core Language}
\label{sec:core_language}

\paragraph*{Syntax}
\label{sec:prog:syntax}

\begin{figure}[hbt]
\parbox{0.5\textwidth}{
	\begin{tabular}{l p{4.5cm}}		
        $P$ ::= & \hfill ($\mathit{program}$) \\
                & $S,t$ \hfill (program shapes and term) \\ 
        \\
        $t$ ::= & \hfill ($\mathit{term}$) \\
		& $t$ $t$ \hfill (application) \\
\ifShort{}{
		& | \textbf{let} $x = t$ \textbf{in} $t$ \hfill (let binding) \\
        & | \textbf{fix} $t$ \hfill (fixed point of $t$) \\
}
        & | \textbf{if} $t$ \textbf{then} $t$ \textbf{else} $t$ \hfill (if-then-else) \\
		& | \textbf{cons} $t$ $t$ \hfill (list constructor) \\
        & | \textbf{null} $t$ \hfill (test for empty list) \\
		& | \textbf{head} $t$ \hfill (head of list) \\
		& | \textbf{tail} $t$ \hfill (tail of list) \\
		& | \textbf{query} $q$ \hfill (query) \\
        & | $t.l$ \hfill (projection) \\
        & | $\{l_i=t_i^{i\in 1\ldots n}\}$ \hfill (record) \\
		& | $x$ \hfill (variable) \\
        & | $\mathit{val}$ \hfill (value) \\
\ifShort{
        \\
        \\
}{}
        \\
        \\
	\end{tabular}
}
\parbox{0.5\textwidth}{
	\begin{tabular}{l p{4.5cm}}	
        $\mathit{val}$ ::= & \hfill ($\mathit{values}$) \\
		& $v$ \hfill (graph node) \\
        & | $\{l_i=\mathit{val}_i^{i\in 1 \ldots n}\}$ \hfill (record) \\
		& | \textbf{nil}[$T$] \hfill (empty list) \\
        & | \textbf{cons} $\mathit{val}$ $\mathit{val}$ \hfill (list constructor) \\
		& | $\lambda(x:T).t$ \hfill (abstraction) \\
        & | true \hfill (true) \\
        & | false \hfill (false) \\
        & \\
		$T$ ::= & \hfill ($\mathit{types}$) \\
		& $s$ \hfill (shape name) \\
		& | $T \rightarrow T$ \hfill (function type) \\
		& | $T$ list \hfill (list type) \\
        & | $\{l_i:T_i^{i\in 1\ldots n}\}$ \hfill (record type) \\
        & | bool \hfill (boolean) \\
        & \\
        $\Gamma$ ::= & \hfill ($\mathit{context}$) \\
        & $\emptyset$ \hfill (empty context) \\
        & | $\Gamma$, $x:T$ \hfill (type binding) 
	\end{tabular}
}
\caption{Abstract syntax of $\lambda_\mathit{SHACL}$.}
\label{fig:prog:syntax}
\end{figure}

\ifShort{}{
\begin{figure}[t!]
\begin{align*}
    \textbf{letrec}~x:T_1=t_1~\textbf{in}~t_2~& \myeq 
    ~\textbf{let}~x =~\textbf{fix}~(\lambda x:T_1.t_1)~\textbf{in}~t_2 \\
\end{align*}
\caption{Syntactical abbreviations.}
\label{fig:programmingAbbrev}
\end{figure}
}
Our core language (Fig.~\ref{fig:prog:syntax}) is a simply typed call-by-value
$\lambda$-calculus. A program is a pair consisting of shapes written for the
program $S$ and a term. Terms ($t$) include%
\ifShort{%
\footnote{Since they show no interesting effects, let statements and a fixpoint operator allowing for recursion, 
    e.g., as necessary to define a \texttt{map} function are omitted. They are
    contained in the technical report submitted with the paper.}}{} 
function application\ifShort{}{%
, \textbf{let}-bindings, a fixed point operator for recursion}
and if-then-else expressions.  
Constructs for lists are included in the language: \textbf{cons}, \textbf{nil},
\textbf{null}, \textbf{head} and \textbf{tail}. Specific to our language is
a querying construct for querying an RDF graph with PCQs.  To avoid confusion
between PCQ query variables and program variables, we refer to the variables of
a query always with the symbol $l$ as they are treated as labels in the program.
We assume labels to be either simple user-defined labels as commonly used in
records, query variables or property paths.  Labels are used for projection. In
case of a projection for a record, the value associated with label is selected.
When evaluating queries, evaluation rules turn query results into lists of
records whereas answer variables are used as record labels. Lastly, in case of
a projection for a graph node, the label is interpreted as a property path and
the graph is traversed accordingly.  Even though not explicitly mentioned in the
syntax, we sometimes add parenthesis to terms for clarification.
Values ($\mathit{val}$)
include graph nodes, record values, nil and cons to represent lists,
$\lambda$-abstractions and the two boolean values true and false.
$\lambda$-abstractions indicate the type of their variable explicitly. 

Types ($T$) include shape names ($s$) as well as type constructors for
function ($T \rightarrow T$), list ($T~\text{list}$) and record types ($\{l_i
: T_i^{i \in 1\ldots n} \}$). We assume primitive data types such as integers
and strings, but omit routine details. To illustrate them, we include booleans
in our syntax.
As common in simply typed $\lambda$-calculi, we also require a context $\Gamma$
for storing type bindings for $\lambda$-abstractions. 

\ifShort{}{
Based on the language, a \textbf{letrec} symbol can be defined (see
Fig.~\ref{fig:programmingAbbrev}). As we lack polymorphism, we cannot define
a general \textbf{map} function. However, we can define a specialized
\textbf{map} function for a record with a single label $x$ of type
$s_\mathit{Student}$ and integer:
\begin{align*}
\textbf{letrec}~\text{map}:
    (\{x : s_\mathit{Student}\} \rightarrow \text{int}) 
    \rightarrow (\{x : s_\mathit{Student}\}~\text{list} 
    \rightarrow \text{int}~\text{list}) = \\  
\lambda (f:\{x : s_\mathit{Student}\} \rightarrow \text{int}) .~
    \lambda (r:\{x : s_\mathit{Student}\}~\text{list}) .~ \\
    \textbf{if}~(\textbf{null}~r)~
    \textbf{then}~\textbf{nil}[\text{int}] \hspace{2.05cm} \\
        ~\textbf{else}~\textbf{cons}~(f~(\textbf{head}~r))~
        (\text{map}~(\textbf{tail}~r))
\end{align*}
}

\ifShort{
As an example, remember the program in Lst.~\ref{lst:intro_age}
which queried for all instances of \texttt{Student}. Assuming that \texttt{map}
is defined using basic recursion, the program can be expressed as}
{The specialized map function can then be used to express the program from
Lst.~\ref{lst:intro_age} in our syntax:}
\[
    \textbf{map}~(\lambda (y:\{x : s_\mathit{Student}\}) . y.x.\texttt{age})~
        (\textbf{query}~(x_1)\leftarrow x~\texttt{type}~\texttt{Student})
\]
In this program, the function ($\lambda$-abstraction) has one variable $y$ whose
type is a record. The record consists of a single label $x$, representing the answer
variable of the query. The type of $x$ is the shape $s_\mathit{Student}$. The
term $y.x$ in the body of the function constitutes an access to the record
label. Accessing the \texttt{age} in the next step constitutes a projection that
traverses the graph. Type-checking rightfully rejects this program as nodes
conforming to $s_\mathit{Student}$ may not have a $\texttt{age}$ relation.  
%
%

\paragraph*{Semantics}
\label{sec:prog:semantics}

The operational semantics is defined using a reduction relation, which extends
the standard ones. As types do not influence run-time behavior, shapes do not
occur in the evaluation rules. However, we define the reduction rules with
respect to an RDF graph $G$. Reduction of lists, records and other routine terms
bear no significant differences from reduction rules as, e.g., defined
in~\cite{Pierce:2002:TPL:509043} 
\ifShort{(c.f. Fig~\ref{fig:sem:standard}, reduction rules for lists are only
        contained in the technical report)}
    {(c.f. Fig.~\ref{fig:sem:standard} and Fig.~\ref{fig:sem:lists})}. 
\begin{figure}[ht!]
    \begin{boxedminipage}{1.0\textwidth}
        \subfile{./sections/semantics/standard.tex}
    \end{boxedminipage}
    \caption{Reduction rules of $\lambda_\mathit{SHACL}$.}
    \label{fig:sem:standard}
\end{figure}
\ifShort{}
    {
\begin{figure}[ht!]
    \begin{boxedminipage}{1.0\textwidth}
        \subfile{./sections/semantics/lists.tex}
    \end{boxedminipage}
    \caption{Reduction rules for lists of a $\lambda$-calculus.}
    \label{fig:sem:lists}
\end{figure}
}
Reduction rules for queries and node projections are summarized by rules E-QUERY
and E-PROJNODE in Fig.~\ref{fig:sem:standard}.  A term representing a query can be directly evaluated
to a list of records.  Query evaluation $\llbracket q \rrbracket_G$ returns
a list of mappings. As in other approaches (e.g.,~\cite{comega}), each query
result becomes a record of the list. For each record, labels are created for
each variable whereas the value of the record is the value provided by the
mapping. A projection on a given graph node is evaluated as a query by turning
the property path expression $l$ into a query pattern. However, instead of
a record a plain list of graph nodes is returned.  


Any term $t$ which cannot be reduced any further (i.e.\ no rule applies to the
term anymore) is said to be in \emph{normal form}. When evaluation is
successful, then the term has been reduced to a value $val$. Any term that is in
normal form but not a value is said to be stuck.  As
usual~\cite{Pierce:2002:TPL:509043}, we use ``stuckness'' as a simple notion of
a run-time error. 

\end{document}

\section{Type system}
\label{sec:type_system}

The most distinguishing feature of the type system is the addition of shape
names as types in the language. As each shape name requires a proper definition,
our typing relation ``:'' is defined with respect to a set of shapes. Likewise,
a typing context $\Gamma$ is required to store type bindings for
$\lambda$-abstractions. Since certain constructs such as queries create new
shapes during the type checking process, the typing relation does not only
assign a type to a term but also a set of newly created shapes which in turn may
contain definitions of shape names that are being used as types. 

\ifShort{

For the typing rules, we require the definition function $\mathit{lub}$ that
computes the least upper bound of two types. The exact definition can be found
in the technical report. Intuitively, in case of two shapes $s_1$
and $s_2$, we rely on disjunction $s_1 \lor s_2$ as a least upper bound.

}
    {

\paragraph*{Least upper bound} For a few constructs, e.g., if-then-else
expressions, require the least upper bound of two types $T_1$ and $T_2$ has to
be constructed through an operator $\mathit{lub}$ (see Fig.~\ref{fig:sem:lub}).
In case of primitive types such as bool, the two types must simply be equal. In
case of two shapes $s_1$ and $s_2$, computing the least upper bound constructs
a new shape $s_\mathit{lub}$ which uses the disjunction of the two shapes as its
constraint ($s_1 \lor s_2$). This requires a new shape name for which we assume
a function $\mathit{genName}$. As a new shape is constructed, $\mathit{lub}$
does not only return a type, but also a set of shapes. The remaining cases are
standard. For lists ($T_1~\text{list}$ and $T_2~\text{list}$), the least upper
bounds of the base types $\mathit{lub}(T_1,T_2,S)$ is constructed. Likewise, for
two functions $T_{11} \rightarrow T_{12}$ and $T_{21} \rightarrow T_{22}$, the
greatest lower bound  $\mathit{glb}$ of the argument types $T_{11}$ and $T_{21}$
(``contra-variance'') as well as the least upper bound of $T_{21}$ and $T_{22}$
(``co-variance'') are computed.  The greatest lower bound of two types is
defined analogously. In case of shapes, conjunction is used. 

\begin{figure}[h!t]
    \begin{boxedminipage}{1\textwidth}
        \subfile{sections/semantics/lub.tex}
    \end{boxedminipage}
    \caption{Least upper bound of two types.}
    \label{fig:sem:lub}
\end{figure}


}

\paragraph*{Typing rules} 
\label{par:typing}

The typing rules for constructs unrelated to querying are mainly the standard
ones as common in simply typed $\lambda$-calculi, except all rules are defined
with respect to a set of shapes and return a set of newly created shapes (see
Fig.~\ref{fig:sem:type_system}). Basic rules, such as for boolean values (rules
T-TRUE and T-FALSE) simply return empty sets of shapes as they do not create new
shapes.  Several rules take possible extensions of the set of shapes into
account. E.g., rule T-PROGRAM takes the set of shapes as defined by the program
$S_P$ and the pre-defined set of shapes $S$ and uses the union of both to
analyze the term $t$. 

\begin{figure}[h!t]
    \begin{boxedminipage}{1\textwidth}
        \subfile{sections/semantics/type-rules.tex}
    \end{boxedminipage}
    \caption{Typing rules for $\lambda_\mathit{SHACL}$.}
    \label{fig:sem:type_system}
\end{figure}


New shapes are mainly created when either the least upper bound judgement is
used or one of the two query expressions (either \textbf{query} or projections)
are used (see rules T-QUERY and T-NPROJ in Fig.~\ref{fig:sem:type_system}). In
case of a \textbf{query} statement (rule T-QUERY), the shape inference rules as
described in Section~\ref{sec:shape_inference_revised} are being used to
construct the set $S_q$ which is being returned as newly created shapes. The
actual type of a query then comprises a list of records. Each record contains
one label per answer variable whereas the type of each label is the respective
shape name for the query variable. Likewise, projections on graph nodes
(T-NODEPROJ) create a new shape name $s'$ using a function $\mathit{genName}$
based on the old shape name $s$ with the appropriate constraint $\geq_1
l^{-}.s$. The newly created definition is returned as a set with the actual type
of the expression being $s~\text{list}$.

\paragraph*{Subtyping}
\label{par:subtyping}

Subtyping rules are summarized in Fig.~\ref{fig:sem:subtyping}. We rely on
a standard subtyping relation. A term $t$ of type $T_1$ is also of type
$T_2$, if $T_1 <: T_2$ is true (T-SUB). Any type is always a subtype of itself
(S-RELF). If $T_1$ is a subtype of $T_2$ and $T_2$ is a subtype of $T_3$, then
$T_1$ is also a subtype of $T_3$ (S-TRANS). Subtyping for lists and functions is
reduced to subtyping checks for their associated types. A list $T_1~\text{list}$
is a subtype of $T_2~\text{list}$ if $T_1$ is a subtype of $T_2$ (S-LIST).
Function types are in a subtyping relation (S-FUNC) if their domains are in
a flipped subtyping relationship (``contra-variance'') and their co-domains are
in a subtyping relationship (``co-variance''). Record type is a subtype of
another record if 1) it has the the same plus more fields (S-RCDWIDTH), 2) it is
a permutation of the supertype (S-RCDPERM) and 3) if the types of the fields are
in a subtype relation (S-RCDDEPTH).

Subtyping relations between two shapes $s_1$ and $s_2$ are defined via faithful
assignments. An assignment $\sigma : N_G \rightarrow 2^{N_S}$ is a function that
assigns shape names to graph nodes. We require the opposite direction---a
function $\sigma_\mathit{inv}$ assigning nodes to shapes. 
\begin{definition}[Inverse assignments]
Let $G$ be an RDF data graph, $S$ a set of shapes and $\sigma^{G,S}$ a faithful
assignment for $G$ and $S$. Then $\sigma_\mathit{inv}^{G,S}$ is a total function
$\sigma_\mathit{inv}^{G,S} : N_S \rightarrow 2^{N_G}$ mapping shape names to
subsets of $N_G$ such that for all graph nodes $v \in N_G$ and all shape names
$s \in N_S$: $s\in \sigma^{G,S}(v)~\text{iff}~v \in \sigma^{G,S}_\mathit{inv}(s)$ 
\end{definition}
For a given set of shapes $S$, two shapes $s_1$ and $s_2$ are in a subtyping
relation if, for all possible RDF graphs $G \in \mathcal{G}$ and all faithful
assignments $\Sigma^{G,S}$ for $S$ and $G$, it holds that
$\sigma^\mathit{inv}_{G,S}(s_1) \subseteq \sigma^\mathit{inv}_{G,S}(s_2)$
(S-SHAPE). That is, the sets of nodes conforming to the two shapes are in
a subset relation for all possible RDF graphs conform to the set of shapes.

\begin{figure}[h!t]
    \begin{boxedminipage}{1\textwidth}
        \subfile{sections/semantics/subtyping.tex}
    \end{boxedminipage}
    \caption{Subtyping rules.}
    \label{fig:sem:subtyping}
\end{figure}

\paragraph*{Algorithmic subtyping} Algorithmic solutions to standard subtyping
rules such used in Fig.~\ref{fig:sem:subtyping} are, e.g., described
by~\cite{Pierce:2002:TPL:509043}. In the case of subtyping for shapes,
algorithmic approaches similar to subsumption checking in description
logics~\cite{dlhandbook} can be employed. That is, $s_1$ must be a subtype of
$s_2$ if it can be shown that no graph exists that contains a node $v$ for
which $s_1 \in \sigma^{G,S}(v)$ but $s_2 \not\in \sigma^{G,S}(v)$.  As of now,
we compare constraint sets which is sound but incomplete. We don't know whether
a complete algorithm exists, although we plan to investigate a transformation
into a description logic based reasoning problem.

\begin{figure}[tb]
    \begin{boxedminipage}{1\textwidth}
        \subfile{sections/semantics/type-elaboration.tex}
    \end{boxedminipage}
    \caption{Type system with type elaboration (excerpt).}
    \label{fig:sem:type-elaboration}
\end{figure}

\paragraph*{Type elaboration} Types do not play any role during the evaluation
of terms. They are only used during the type checking process. This is by
design, as run-time type checks incur overhead and should be avoided, in
particular if the type check is computationally expensive.  However, the
evaluation relation only evaluates terms of the form $v.l$ (node projections)
into lists of graph nodes (c.f.~rule E-PROJNODE of Fig.~\ref{fig:sem:standard} and
T-NPROJ of Fig.~\ref{fig:sem:type_system}), even though a shape may
hint that there is only one successor (e.g., \texttt{studiesAt} of shape
$s_\mathit{Student}$). As the evaluation rules have no information about types,
the type system must annotate or transform terms such that they can be treated
differently during run-time. This process is called \emph{type
elaboration}~\cite{Pierce:2002:TPL:509043}.
The typing relation ``:'' then takes a set of shapes $S$ and a typing context
$\Gamma$ and returns a term $t$, a type $T$ and a set of newly introduced shapes
$S'$. This is exemplified by the rules in Fig.~\ref{fig:sem:type-elaboration}.
Most rules simply return the term without modifications (e.g., rule T-HEAD).
However, in case of node projections where it can be shown that there is only
a single successor, a $\textbf{head}$ is automatically added to the term (rule
T-NPROJ-1). Otherwise, the term is not modified (rule T-NPROJ-2).

%

\end{document}

\section{Type Soundness}
\label{sec:type_soundness}

A term $t$ is said to be well-typed if the type system assigns a type. We show
the soundness of the $\lambda_\mathit{SHACL}$ type system by proving that
a well-typed term does not get stuck during evaluation. As with other languages,
there are exceptions to this rule, e.g., down-casting in object-oriented
languages, c.f.~\cite{featherweightJava}. For $\lambda_\mathit{SHACL}$, this
exception concerns lists.  
We show that if a program is well-typed, then the only way it can get stuck is by
reaching a point where it tries to compute \textbf{head nil} or \textbf{tail
nil}. Furthermore, terms must be closed, meaning that all program variables are
bound by function abstractions~\cite{Pierce:2002:TPL:509043}. We proceed in two
steps, by showing that a well-typed term is either a value or it can take a step
(progress) and by showing that if that term takes a step, the result is also
well-typed (preservation). 

\begin{lemma}[Canonical Forms Lemma]\label{lemma:canonicalForms}
    Let $\mathit{val}$ be a well-typed value. Then the following observations
    can be made:
    \begin{enumerate}
        \item If $\mathit{val}$ is a value of type $s$, then $\mathit{val}$ is
            of the form $v$.
        \item If $\mathit{val}$ is value of type $T_1 \rightarrow T_2$, then
            $\mathit{val}$ is of the form $\lambda (x:T_1).t_2$.
        \item If $\mathit{val}$ is a value of type $T~\text{list}$, then
            $\mathit{val}$ is either of the form $\textbf{cons}~\mathit{val}
            \ldots$ or $\textbf{nil}$. 
        \item If $\mathit{val}$ is a value of type $\{l_i : T_i^{i \in 1 \ldots
            n}\}$, then $\mathit{val}$ is of the form $\{l_i
                = \mathit{val}_i^{i\in 1 \ldots n }\}$.
        \item If $\mathit{val}$ is a value of type $\text{bool}$, then
            $\mathit{val}$ is  either of the form $\text{true}$ or
            $\text{false}$.  
    \end{enumerate}
\end{lemma}

Given Lemma~\ref{lemma:canonicalForms}, we can show that a well-typed term is
either a value or it can take a step. 

\begin{theorem}[Progress]\label{th:progress}
    Let $t$ be a closed, well-typed term. If $t$ is not a value, then there
    exists a term $t'$ such that $t \rightarrow t'$. If $S,\Gamma \vdash t : T,
    S'$, then $t$ is either a value, a term containing the forms \textbf{head
    nil} or \textbf{tail nil}, or there is some $t'$ with $t \rightarrow t'$. 
\end{theorem}

\ifShort{

\begin{proof}[Proof (Sketch)]  
    The theorem can be shown by induction on the derivation of $S,\Gamma \vdash
    t : T, S$. 
   %
    Queries ($t=\textbf{query}~q$) are straightforward as no sub-term exists.
    For node projections ($t_1.l$ with the type of $t_1$ being a shape name),
    Lemma~\ref{lemma:canonicalForms} tells us that it must ultimately reduce to
    a graph node. In that case rule E-PROJNODE applies.  The full proof can be
    found in the tech report.
\end{proof}

}
    {

\begin{proof}
    By induction on the derivation of $S,\Gamma \vdash t : T, S$. 
    \begin{description}

        \item[T-APP] 
            $t = t_1 t_2$,
            $S,\Gamma \vdash t_1 : T_{11} \rightarrow T_{12},S_1$,
            $S,\Gamma \vdash t_2 : T_{11}, S_2$.
            By hypothesis, $t_1$ and $t_2$ are either values or they can take
            a step. If they can take a step, rules E-APP1 or E-APP2 apply. If
            both are values, then by the canonical forms lemma
            (Lemma~\ref{lemma:canonicalForms}), $t_1 = \lambda
            (x:T_{11}).t_{11}$ and rule E-APPABS applies.

        \item[T-LET] 
            $t =\textbf{let}~x=t_1~\textbf{in}~t_2$,
            $S,\Gamma \vdash t_1 : T_1,S_1$,
            $S\cup S_1, (\Gamma, x:T_1) \vdash t_2 : T_2,S_2$.
            By hypothesis, $t_1$ is either a value or it can make a step. If it
            can, then rule E-LET applies. If it is a value, then rule (E-LETV)
            applies.

         \item[T-FIX] 
            $t = \textbf{fix}~t_1$,
            $S,\Gamma\vdash t:T_1,S_1$. 
            $S,\Gamma \vdash t_1:T_1\rightarrow T_1,S_1$,
            By induction hypothesis, $t_1$ is either a value or it can take
            a step.  If it can take a step, rule E-FIX applies. If its a value,
            by the canonical forms lemma (Lemma~\ref{lemma:canonicalForms}), $t_1
            = \lambda(x:T_1).t_2$.  Therefore, rule E-FIXBETA applies. 

        \item[T-IF]  
            $t = \textbf{if}~t_1~\textbf{then}~t_2~\textbf{else}~t_3$,
            $S,\Gamma \vdash t_1 : \text{bool}, S_1$. 
            By hypothesis, $t_1$ is a value or it can take a step. If it can
            take a step, rule E-IF applies. If it is a value, then by
            Lemma~\ref{lemma:canonicalForms}, either $t_1 = \text{true}$ or $t_1
            = \text{false}$. In this case, either rules E-IF-TRUE or E-IF-FALSE
            apply. 

        \item[T-NIL] Immediate, since \textbf{nil} is a value.
        
        \item[T-CONS]
            $t = \textbf{cons}~t_1~t_2$,
            $S,\Gamma \vdash t_1 : T_1,S_1$,
            $S,\Gamma \vdash t_2 : T_1~\text{list},S_2$.
            By hypothesis, $t_1$ and $t_2$ are either values or they can take
            a step. If they can take a step, then rules E-CONS1 or E-CONS2
            apply. If both $t_1$ and $t_2$ are values, then $t$ is also a value.

        \item[T-ABS] Immediate, since $\lambda (x:T).t_1$ is value. 
        \item[T-VAR] Impossible since we're only looking at closed terms. 
        \item[T-TRUE] Immediate, since $\text{true}$ is a value. 
        \item[T-FALSE] Immediate, since $\text{false}$ is a value. 
        
        \item[T-NULL]
            $t = \textbf{null}~t_1$,
            $S,\Gamma \vdash t_1 : T_1~\text{list},S_1$.
            By hypothesis, $t_1$ is a value or it can take a step. If it can
            take a step, then rule E-NULL applies. If it is a value, then by
            Lemma~\ref{lemma:canonicalForms}, $t = \textbf{nil}$ or $t
            = \textbf{cons} \mathit{val}_1 \ldots$. Then either rule E-NULL-TRUE
            or E-NULL-FALSE apply. 
        
        \item[T-HEAD]
            $t = \textbf{head}~t_1$,
            $S,\Gamma \vdash t_1 : T_1~\text{list}, S_1$. 
            By hypothesis, $t_1$ is either a value or it can take a step. If it
            can take a step, rule E-HEAD applies. If it is a value, then by
            Lemma~\ref{lemma:canonicalForms}, either $t = \textbf{nil}$ or $t
            = \textbf{cons}~\mathit{val}_1 \ldots$. Then either rule E-HEADV
            applies or the term is in the accepted normal form $t = \textbf{head
            nil}$. 

        \item[T-TAIL]
            $t = \textbf{tail}~t_1$,
            $S,\Gamma \vdash t_1 : T_1~\text{list},S_1$.
            By hypothesis, $t_1$ is either a value or it can take a step. If it
            can take a step, then rule E-TAIL applies. If it is a value, then by
            Lemma~\ref{lemma:canonicalForms}, either $t = \textbf{nil}$ or $t
            = \textbf{cons}~\mathit{val}_1 \ldots$. Then either rule E-TAILV
            applies or the term is in the accepted normal form $t = \textbf{tail
            nil}$. 

        \item[T-RCD]
            $t = \{l_i:T_i^{i\in 1 \ldots n}\}$,
            $\text{for each } i~S,\Gamma \vdash t_i : T_i, S_i$.
            By induction hypothesis, each $t_i$ is either a value or it can take
            a step. If one can take a step, then rule E-RCD applies. If each
            $t_i$ is a value, then $t$ is also a value. 

        \item[T-RCDPROJ]
            $t = t_1.l_i$,
            $S,\Gamma \vdash t_1 : \{l_i : T_i^{i\in 1 \ldots n}\}$.
            By hypothesis, $t_1$ is either a value or it can take a step. If it
            can take a step, then rule E-PROJ applies. If it is a value, then by
            Lemma~\ref{lemma:canonicalForms}, then $t = \{l_i : \mathit{val}_i^{i \in
            1 \ldots n}\}$ and rule E-PROJRCD applies. 

        \item[T-QUERY] Immediate since rule E-QUERY applies. 

        \item[T-NPROJ]
            $t = t_1.l$,
            $S,\Gamma \vdash t_1 : s,S_1$.
            By induction hypothesis, $t_1$ is either a value or it can take
            a step. If it can take a step, then rule E-PROJ applies. If it is
            a value, then by Lemma~\ref{lemma:canonicalForms}, $t = v$ and rule
            E-PROJNODE applies. 

        \item[T-SUB] Results follow from induction hypothesis. 
    \end{description}
\end{proof}

}

\ifShort{Given that a well-typed term can take a step, we now need to show that
    taking a step according to the evaluation rules preserves the type.}
    { 
    For proving preservation, an additional Lemma is required stating that
    substitution, as for example used when evaluating \textbf{let}-statements or
    function applications, preserves the type.  
    
    \begin{lemma}[Substitution]\label{lemma:substitution}
        If $S, (\Gamma,x:T_2) \vdash t : T_1,S_1$ and $S,\Gamma \vdash t_2 : T_2,
        S_2$, then $S,\Gamma \vdash \lbrack x \mapsto t_2 \rbrack t_1 : T_1, S'$.  
    \end{lemma}

    \begin{proof}     
        Substitution in our case does not differ from standard approaches, e.g.,
        as described by~\cite{Pierce:2002:TPL:509043}. Therefore, the proof is
        omitted.  
    \end{proof}

    We can now show that if a term takes a step by the evaluation rules, its type is
    preserved. 
}

\begin{theorem}[Preservation] \label{th:preservation}
Let $t$ be a term and $T$ a type. If $S,\Gamma \vdash t : T, S'$ and $t
\rightarrow t'$, then $S,\Gamma \vdash t':T,S'$.
\end{theorem}

\ifShort{

\begin{proof}[Proof (Sketch)]
As with progress, the proof is an induction over the typing relation $S,\Gamma
\vdash t : T,S'$. For each term, possible ways of reducing it are distinguished and
it is shown that in each case the type does not change. 
%
For queries, this is immediate. In case of node projections, $t_1$ either
took a step, in which case the typing rule applies again, or it is a graph node
$v$ with type $s$. Each $v'$ which is reached via the node projection conforms
to the newly created shape $s'$ with its constraint $\geq_1 l^{-}.s$. Therefore,
the type is also preserved.
\end{proof}

}
    {

\begin{proof}
By induction of the derivation of $S,\Gamma \vdash t : T,S'$.
    \begin{description}

        \item[T-APP] 
            $t = t_1 t_2$,
            $S,\Gamma \vdash t_1 : T_{11} \rightarrow T_{12},S_1$,
            $S,\Gamma \vdash t_2 : T_{11}, S_2$,
            $S,\Gamma \vdash t_1 t_2 : T_{12}, S_1 \cup S_2$. 
            There are three rules by which $t'$ can be derived: E-APP1, E-APP2
            and E-APPABS. 
            \begin{enumerate}
                \item $t' = t_1' t_2$ By induction hypothesis, $t_1 \rightarrow
                    t'_1$ preserves the type. Therefore, by rule T-APP, $t'
                    : T_{12}$. 
                \item $t' = \mathit{val}_1 t'_2$. Same as first case. 
                \item $t' = \lbrack x \mapsto \mathit{val}_2 \rbrack t_{12}$. By
                    Lemma~\ref{lemma:substitution}, the type is preserved.
                    Therefore $t' : T_{12}$. 
            \end{enumerate}

        \item[T-LET] 
            $t =\textbf{let}~x=t_1~\textbf{in}~t_2$,
            $S,\Gamma \vdash t_1 : T_1,S_1$,
            $S \cup S_1, (\Gamma, x:T_1) \vdash t_2 : T_2,S_2$.
            $S,\Gamma \vdash t : T_2,S_1 \cup S_2$,
            There are two ways $t$ can be reduced: E-LET and E-LETV.
            \begin{enumerate}
                \item $t' = \textbf{let}~x = t'_1~\textbf{in}~t_2$. By induction
                    hypothesis, $t_1 \rightarrow t'_1$ preserves the type. Then
                    by rule T-LET, $t' : T_2, S_1\cup S_2$. 
                \item $t' = \lbrack x \mapsto \mathit{val}_1 \rbrack t_2$. By
                    Lemma~\ref{lemma:substitution}, the type is preserved,
                    therefore $t' : T_2$. 
            \end{enumerate}

        \item[T-FIX] 
            $t = \textbf{fix}~t_1$,
            $S, \Gamma\vdash t_1:T_1\rightarrow T_1,S_1$,
            $S, \Gamma\vdash t:T_1, S_1$. 
            There are two rules by which $t$ can be reduced: E-FIX and E-FIXBETA.  
            \begin{enumerate}             
                \item $t'=\textbf{fix}~t'_1$. By induction hypothesis, $t_1
                    \rightarrow t'_1$ preserves the type.  Then, by T-FIX,
                    $t':T_1$.  
     
                \item $t'=\lbrack x \mapsto \textbf{fix}~(\lambda (x:T_1).t_2)\rbrack t_2$. 
                    By Lemma~\ref{lemma:substitution}, the type is preserved,
                    therefore $t' : T_1$.
            \end{enumerate}

        \item[T-IF]  
            $t = \textbf{if}~t_1~\textbf{then}~t_2~\textbf{else}~t_3$,
            $S,\Gamma \vdash t_1 : \text{bool}, S_1$,
            $S,\Gamma \vdash t : T_\mathit{lub}, S_1 \cup S_2 \cup S_3$.
            There are three rules by which $t'$ can be derived: E-IF, E-IF-TRUE
            and E-IF-FALSE. 
            \begin{enumerate}
                \item $t'=\textbf{if}~t'_1~\textbf{then}~t_2~\textbf{else}~t_3$.
                    By hypothesis, $t_1 \rightarrow t'_1$ preserves the type.
                    Therefore, by rule T-IF, $t : T_\mathit{lub}$.
                \item $t_1 = \text{true}$, $t' = t_2$. By the construction of
                    $T_\mathit{lub}$, it must be true that $t_2
                    <: T_\mathit{lub}$. Therefore, $t' : T_\mathit{lub}$.
                \item $t_1 = \text{false}$, $t' = t_3$. Same as second case. 
            \end{enumerate}

        \item[T-NIL] Vacuously fulfilled, since \textbf{nil} is a value.
        
        \item[T-CONS]
            $t = \textbf{cons}~t_1~t_2$,
            $S,\Gamma \vdash t_1 : T_1,S_1$,
            $S,\Gamma \vdash t_2 : T_1~\text{list},S_2$,
            $S, \Gamma \vdash t : T_1~\text{list},S_1 \cup S_2$. 
            There are two rules by which $t'$ can be derived: E-CONS1 and
            E-CONS2. 
            \begin{enumerate}
                \item $t' = \textbf{cons}~t'_1 t_2$. By hypothesis, $t_1
                    \rightarrow t'_1$ preserves the type. Therefore, by rule
                    T-CONS, $t' : T_1~\text{list}$. 
                \item $t' = \textbf{cons}~\mathit{val}_1 t'_2$. Same as first
                    case.
            \end{enumerate}

        \item[T-ABS] Vacuously fulfilled, since $\lambda (x:T).t_1$ is value. 
        \item[T-VAR] Cannot happen. 
        \item[T-TRUE] Vacuously fulfilled, since $\text{true}$ is a value. 
        \item[T-FALSE] Vacuously fulfilled, since $\text{false}$ is a value. 
        
        \item[T-NULL]
            $t = \textbf{null}~t_1$,
            $S, \Gamma \vdash t_1 : T_1~\text{list}, S_1$,
            $S, \Gamma \vdash t : \text{bool}, S_1$.
            There are three rules by which $t'$ can be derived: E-NULL,
            E-NULL-TRUE and E-NULL-FALSE. 
            \begin{description}
                \item $t' = \textbf{null}~t'_1$. By hypothesis, $t_1 \rightarrow
                    t'_1$ preserves the type. Therefore, by rule T-NULL, $t'
                    : \text{bool}$. 
                \item $t'_1 = \textbf{nil}$, $t' = \text{true}$. By rule T-TRUE,
                    $t' : \text{bool}$. 
                \item $t'_1 = \textbf{cons}~\mathit{val}_1 \ldots$, $t'
                    = \text{false}$. By rule T-FALSE, $t' : \text{bool}$. 
            \end{description}

        \item[T-HEAD]
            $t = \textbf{head}~t_1$,
            $S,\Gamma \vdash t_1 : T_1~\text{list}, S_1$. 
            $S, \Gamma \vdash t : T_1, S_1$. 
            There are two rules by which $t'$ can be derived: E-HEAD and
            E-HEADV. 
            \begin{enumerate}
                \item $t' = \textbf{head}~t'_1$. By hypothesis, $t_1 \rightarrow
                    t'_1$ preserves the type. Therefore, by rule T-HEAD, $t'
                    : T_1$. 
                \item $t_1 = \textbf{cons}~\mathit{val}_1 \ldots$, $t'
                    = \mathit{val}_1$. Due to rules T-CONS and T-HEAD,
                    $\mathit{val}_1$ must have type $T_1$. Therefore, $t'
                    : T_1$. 
            \end{enumerate}

        \item[T-TAIL]
            $t = \textbf{tail}~t_1$,
            $S,\Gamma \vdash t_1 : T_1~\text{list},S_1$,
            $S, \Gamma \vdash \textbf{tail}~t_1 : T_1~\text{list}, S_1$. 
            There are two rules by which $t'$ can be derived: E-TAIL and
            E-TAILV. 
            \begin{enumerate}
                \item $t' = \textbf{tail}~t'_1$. By hypothesis, $t_1 \rightarrow
                    t'_1$ preserves the type. Therefore, by rule T-TAIL, $t'
                    : T_1~\text{list}$.
                \item $t_1 = \textbf{cons}~\mathit{val}_1~\mathit{val}_2$, $t'
                    = \mathit{val}_2$. Due to rules T-CONS and T-TAIL,
                    $\mathit{val}_2$ must have type $T_1~\text{list}$,
                    Therefore, $t' : T_1~\text{list}$. 
            \end{enumerate}

        \item[T-RCD]
            $t = \{l_i=t_i^{i\in 1 \ldots n}\}$,
            $\text{for each } i~S,\Gamma \vdash t_i : T_i, S_i$,
            $S,\Gamma \vdash t : \{l_i : T_i^{i \in 1 \ldots n}\}, \bigcup_i=1^n
            S_i$.
            $t'$ can only be derived be rule E-RCD in which $t_i \rightarrow
            t'_i$. By hypothesis, this preserves the type.  
        
        \item[T-RCDPROJ]
            $t = t_1.l_j$,
            $S,\Gamma \vdash t_1 : \{l_i : T_i^{i\in 1 \ldots n}\}, S_1$,
            $S, \Gamma \vdash t : T_j, S_1$. 
            There are two rules by which $t'$ can be derived: E-PROJ and
            E-PROJRCD. 
            \begin{enumerate}
                \item $t' = t'_1.l_j$. By hypothesis, $t_1 \rightarrow t'_1$
                    preserves the type. Therefore, $t' : T_j$.
                \item $t = \{l_i = \mathit{val}_i^{i\in 1 \ldots n}\}$, $t'
                    = \mathit{val}_j$. Due to rule T-RCD and T-RCDPROJ,
                    $\mathit{val}_j$ must have type $T_j$. Therefore, $t'
                    : T_j$. 
            \end{enumerate}

        \item[T-QUERY] Immediate since rule E-QUERY applies. 

        \item[T-NPROJ]
            $t = t_1.l$,
            $S,\Gamma \vdash t_1 : s,S_1$.
            $S, \Gamma \vdash t : s'~\text{list}, S_1 \cup \{s',\geq_1 l^{-}.s,
            \emptyset\}$.
            There are two rules by which $t'$ can be derived: E-PROJ and
            E-PROJNODE. 
            \begin{enumerate}
                \item $t' = t'_1.l$. By hypothesis, $t_1 \rightarrow t'_1$
                    preserves the type. Therefore, $t' : s'$. 
                \item $t_1 = v$, $t'
                    = \textbf{cons}~\mu_1(x)~\ldots~\textbf{cons}~\mu_n(x)~\textbf{nil}$
                    with $\mu_i \in \llbracket l(v,x) \rrbracket_G$. Each node
                    $\mu_i(x)$ must fulfill the constraint $\geq_1 l^{-}.s$ of
                    shape $s'$ as it would otherwise not be in the query result.
                    Therefore, the type is preserved as $t' : s'~\text{list}$   
            \end{enumerate}

        \item[T-SUB] Results follows from induction hypothesis. 

\end{description}
\end{proof}

}

As a direct consequence of Theorems~\ref{th:progress} and \ref{th:preservation},
a well-typed, closed term does not get stuck during evaluation. 

\end{document}

\section{Related Work}
\label{sec:related_work}

The presented approach is generally related to the validation of RDF as well as
the integration of RDF into programming languages. 
RDF validation has seen an increase in interest. Among them are inference-based
approaches such as~\cite{icOWL1,icOWL2}, in which OWL expressions are used as
integrity constraints by relying on a closed-world assumption. The fact that
constraints are OWL expressions puts these approaches closer to \cite{lambdadl}
than the approach described here. A validation approach that is relatively
similar to SHACL is ShEx~\cite{shex}. ShEx also uses shapes to group
constraints, but removes property path expressions and features well-defined
recursion. We chose SHACL over ShEx due to SHACL being a W3C recommendation.
Due to the similarity between SHACL and ShEx, the integration process for the
latter is very similar. In fact, the definition for recursion used in ShEx even
simplifies some aspects as there is no need for the notion of faithful
assignments. 

In terms of integration of RDF into programming languages, we consider different
approaches. Generic representations, e.g., the OWL API~\cite{owlapi} or
Jena~\cite{jena}, use types on a meta-level (e.g., \emph{Statement}) that do not
allow a static type-checker to verify a program. This leaves correctness
entirely on the hands of the programmer. 
%
Mapping approaches use schematic information of the data
model to create types in the target language. Type checking can offer some
degree of verification. 
An early example of this is OWL2Java~\cite{owl2java}, a more recent one is 
LITEQ~\cite{liteq}. However, mapping
approaches based on ontologies come with their own limitations. OWL relies on a open-world
assumption, in which missing information is treated as incomplete data rather
than constraint violations. As shown in the introduction, structural information
does therefore not necessarily imply the presence of data relationships. This is
problematic for type-checkers as they rely on a closed world.   
%
%
The most powerful approaches create new languages or extend existing ones to accomodate
the specific requirements of the data model. 
Examples include rule-based programming~\cite{rulebasedProgramming} as well as
a transformation and validation language~\cite{shapeExpressions}. However, both
are untyped.
Typed approaches to linked data is provided
by~\cite{rwLinkedData,LDtypeInference}. Zhi\#~\cite{zhiSharp}, an
extension of the C\# language provides an integration for OWL ontologies, albeit
it only considers explicitly given statements. Contrary
to that, \cite{lambdadl,scaspa} provides an integration of OWL ontologies also
considering implicit statements. However, as shown in the introduction,
programmers cannot rely on structural restrictions given by OWL ontologies
whereas SHACL enforces its structural restriction with a closed-world
assumption.  
%
%

\section{Summary and Future Work}
\label{sec:summary}

In this paper, we have presented an approach for type checking programs using SHACL.
We have shown that by using SHACL shapes as types, type safety can be achieved.
This helps in writing less error-prone programs, in particular when facing
evolving RDF graphs. The work can be extended in several directions. 

First, an implementation of the presented approach is highly desirable.
Comparably to~\cite{scaspa}, we plan on implementing the approach in Scala using
compiler plugins that add new compilation phases. Shape names constitute a new
form of types. As shape names are known before compilation, they can be
syntactically integrated using automatically generated type aliases to a base
type. This allows for type checking shape types in a separate compilation phase
that runs after the standard Scala type inference and type checker phases. As
there is little interaction between normal Scala types and shape types, issues
only arise when code converts e.g., literals into standard Scala types. However,
this can be solved through minor code transformations before the type checking
phase. Lastly, transformations based on type elaboration can also run as
a separate phase. As shape types do not influence run-time behavior, compilation
produces standard JVM byte code. However, one noteworthy limitation of using
type aliases to represent shape names is that method overloading based on shape
names is not possible. Resolving this issue requires better integration
techniques which remain as future work.

Second, finding sound and complete methods for deciding shape subsumption is an
interesting problem that requires future research. This is an important step as
it defines practical boundaries in terms of the parts of SHACL that can be used
for type checking. 
Lastly, the supported subset of SPARQL queries is relatively small and should be
extended by missing features such as union of queries or filter expressions.
This raises questions about the parts of SPARQL that can be described with SHACL
shapes. 

%

\bibliographystyle{splncs04}
\bibliography{paper}



\end{document}